\documentclass[5p,times]{elsarticle}
\usepackage{lineno}
\modulolinenumbers[5]
\usepackage{amsmath}
\usepackage{graphicx,caption}
\usepackage{float}
\usepackage{lipsum}
\usepackage{algorithm}
\usepackage{algpseudocode}
\usepackage{natbib}
\usepackage{siunitx}
\usepackage{xcolor}
\usepackage{xparse}
\usepackage{setspace}
\usepackage{soul}
\usepackage{hyperref}

\newcommand{\argmax}{\operatornamewithlimits{argmax}}
\newcommand{\argmin}{\operatornamewithlimits{argmin}}
\newcommand{\norm}[1]{\left\lVert#1\right\rVert}
\begin{document}
	\begin{frontmatter}
		\title{KLT picker: Particle picking using data-driven optimal templates}
		\author[1]{Amitay Eldar\corref{cor1}}
		\ead{amitayeldar@mail.tau.ac.il}
		\author[2]{Boris Landa}
		\ead{boris.landa@yale.edu}
		\author[1]{Yoel Shkolnisky}
		\ead{yoelsh@tauex.tau.ac.il}
		\cortext[cor1]{Corresponding author}

		\address[1]{Department of Applied Mathematics, School of Mathematical Sciences, Tel-Aviv University, Tel-Aviv ,Israel}
		\address[2]{Department of Mathematics, Yale University, 10 Hillhouse Ave, New Haven, USA}

\begin{abstract}
	Particle picking is currently a critical step in the cryo-EM single particle reconstruction pipeline.
	Despite extensive work on this problem, for many data sets it is still challenging, especially for low SNR micrographs.
	We present the KLT (Karhunen Loeve Transform) picker, which is fully automatic and requires as an input only the approximated particle size. In particular, it does not require any manual picking. Our method is designed especially to handle low SNR micrographs. It is based on learning a set of optimal templates through the use of multi-variate statistical analysis via the Karhunen Loeve Transform. We evaluate the KLT picker on publicly available data sets and present high-quality results with minimal manual effort.
\end{abstract}

\end{frontmatter}

	\section{Introduction}
	Single Particle Cryo-Electron Microscopy is a well known method used for 3D reconstruction of macro-molecules from their multiple 2D projections. In this process, a sample containing macro-molecules is rapidly frozen in a thin layer of ice, whereby the locations and orientations of the frozen samples in the ice layer are unknown. An electron beam is transmitted through the frozen samples creating their 2D projections. The resulting 2D micrographs suffer from low signal-to-noise ratio (SNR).
	The images containing the projections are called micrographs and the projections without the noise are called  particles. The micrographs mostly contain three types of regions -- regions of particles with added noise, regions of noise only, and regions of contaminations.
	One of the first steps towards 3D reconstruction is locating the particles on the micrographs, a step called ``particle picking''.
	In order to achieve a high resolution 3D model of the macro-molecule, one needs many thousands of particle images (to overcome the high levels of noise).
	
	Over the last years, many algorithms aimed to automate the particle picking step have been proposed. Roughly speaking, such algorithms can be divided into two main categories: semi-automatic and automatic.
	
	Semi-automatic algorithms require the user to select manually hundreds to thousands of particles which are used in various ways in order to automatically detect more particles. RELION~\citep{scheres2015semi} and SIGNATURE~\citep{chen2007signature} use  manually picked particles as templates and correlate them against the micrographs in order to compute statistical similarity measures. The main idea is that cross-correlation between a template and a particle image with added noise will be larger than cross-correlation with noise only. Another approach, gaining popularity recently, is the use of methods from  the field of deep learning. TOPAZ~\citep{bepler2018positive} uses manually selected particles to train a convolutional neural network, where Xmipp~\citep{sorzano2004xmipp}  uses them to train an ensemble of classifiers. In both cases, the network and the ensemble are later used to detect more particles. Common to all semi-automatic algorithms is the need for a high quality set of manually picked particles. Most of them could also benefit from a collection of images that contain only noise.
	
	Automatic algorithms do not require the user to manually select particles. Nevertheless, they do require some prior knowledge about the macro-molecule. Most of the algorithms require at least an estimate of the particle size (i.e the maximal diameter of the particle) and some of them require the user to adjust all sorts of parameters. The DoG picker~\citep{voss2009dog}  is based on convolving difference of gaussian functions with a micrograph image. This method is suitable for identifying blobs in the image and to sort them by their sizes.
	The APPLE picker~\citep{heimowitz2018apple}  uses cross-correlation between micrograph patches in order to detect areas that are more likely to contain a particle. These areas are then being used to train a support vector machine, which is used to detect particles. Deep learning methods are also gaining popularity in the automatic category. DeepPicker~\citep{wang2016deeppicker}  and fast R-CNN~\citep{xiao2017fast}  use 3D reconstructions of known macro-molecules as labeled data sets to train neural networks  that are later used to pick particles. This procedure leads to a significant speedup of the particle picking step.
	
	In this paper we present the KLT (Karhunen Loeve Transform) picker, an automatic particle picking method in which the only input needed by the user is an estimate of the particle's size. Our method is based on computing for each micrograph a set of data-adaptive templates, which are optimal in the sense that they admit maximal correlation with the particle images. Additionally, they are naturally adapted to coping with the unknown in-plane rotations of the particles, as the absolute value of their correlation with a particle image is rotationlly invariant. The above-mentioned properties of our templates eventually allow to detect particles under low SNR conditions. The estimation of these templates is achieved through the Karhunen Loeve Transform~\citep{maccone2009simple} by utilizing the radial power spectral density (RPSD)~\citep{papoulis1977signal} of the particles. Using these templates gives rise to a score for each patch in the micrograph, based on the well known Likelihood Ratio Test (LRT)~\citep{wasserman2013all}. Specifically, the algorithm partitions each micrograph into overlapping patches, and assigns to each of them a score that reflects the probability that it contains a particle. The threshold probability for detecting particles is set either to the theoretical threshold (explained below), or, the user can adjust the threshold to pick a smaller number of patches with the highest scores  (high probability for the presence of a particle). In addition the threshold probability can be set to pick  patches that are more likely to contain only noise (patches with the lowest scores).
	One of the advantages of our method is that it can complement other semi-automatic algorithms, that require an initial set of training images. For example, the user can decide to pick only the 10 highest and lowest scored patches out of each micrograph and use them instead of an initial manual picking step.
	Nevertheless, the KLT picker is fully capable of preforming the entire particle picking step, as we demonstrate on three different data sets in Section~\ref{sec:expRes}.
	In order to evaluate the picking results, we perform a 3D reconstruction for each data set and compare the resulting gold-standard Fourier Shell Correlation (FSC) resolution~\cite{vanHeel_Schatz} to the published one. The data sets used are Plasmodium Falciparum 80S ribosome~\citep{PMID:24913268}, $ \beta $-Galactosidase~\citep{PMID:25953817} and Synaptic RAG1-RAG2~\citep{PMID:26548953}.
	The code for the KLT picker is available at  \url{https://github.com/amitayeldar/KLTpicker}.
\section{Materials and methods}\label{sec: material}
We next detail the different steps in our particle picking algorithm.
\subsection{Computing data adaptive particle templates}\label{subsec:optimalTemplates}
Let $ a\in\mathbb{R_+} $ be the particle's radius and let $f\in L^2(\mathbb{R}^2)$ be a function representing a centered particle whose energy is concentrated in a disk of radius $a$. We consider $ f $ to be a random function, with some probability distribution capturing the variability of particle images.
For the purpose of particle picking, we seek the set of orthogonal functions that maximize their correlation with the particle images. Specifically,
            \begin{equation}\label{eq:templatesDef}
			\begin{gathered}
			\psi_n  =\operatornamewithlimits{argmax}_{\hat{\psi}\in L^2(\mathbb{R}^2)} \mathbb{E}\left| \left \langle f,\hat{\psi} \right\rangle_{L^2(a\mathbb{D})}\right|^2.\\
			\text{s.t.}\;\;  \norm{\hat{\psi}}_{L^2(a\mathbb{D})}=1,\;\;\left \langle\hat{\psi},\psi_j \right\rangle_{L^2(a\mathbb{D})}=0, \;\;  j< n\in \mathbb{N},
			\end{gathered}
            \end{equation}
			where $ a\mathbb{D} $ is the disk of radius $ a $, $ \left \langle \cdot,\cdot \right\rangle_{{L^2(a\mathbb{D})}} $ denotes the standard inner product on $ {L^2(a\mathbb{D})} $, and $ \norm{\cdot}_{L^2({a\mathbb{D}})} = \sqrt{\left \langle \cdot,\cdot \right\rangle_{L^2{(a\mathbb{D})}}}$ is the appropriate norm.
			In words, $\psi_1$ is the function which maximizes the correlation with the particle function $f$, $\psi_2$ is the function which maximizes the correlation with $f$ while being orthogonal to $\psi_1$, and so on. These functions are the results of the Karhunen Loeve Transform (KLT)~\citep{maccone2009simple} of the particle function $f$.
			We denote the maximal expected value attained in the right-hand side of~\eqref{eq:templatesDef} by $ \lambda_n $, i.e.
			\begin{align}\label{eq:templatesEqLambda}
				\lambda_n  := \mathbb{E}\left| \left \langle f,\psi_n \right\rangle_{L^2(a\mathbb{D})}\right|^2.
			\end{align}
			$\lambda_n$ describes the importance of every function $\psi_n$ in serving as a template for particle picking, as correlating $\psi_n$ corresponding to a large $\lambda_n$ against a noisy particle image is expected to produce a larger value (on average) than correlating it against pure noise (which would produce the same correlation regardless of the template used).
			
			Solving~\eqref{eq:templatesDef}  (under assumption detailed below ) requires the Radial Power Spectral Density (RPSD)~\citep{papoulis1977signal} of $ f $. In the next section, we present a method to estimate the RPSD under the assumptions that $f$ is band-limited and is generated from a wide sense stationary random process.
			Under these assumptions, we show in~\ref{app:templates derivation} that the solutions to~\eqref{eq:templatesDef}, in polar coordinates, are of the form
			\begin{equation} \label{eq:psi_hat steerable form}
			\psi_{m,k}(r,\theta) = \frac{1}{\sqrt{2\pi}}R_{m,k}(r)e^{im\theta}, \;\;\;\:m\in\mathbb{Z},\;k\in \mathbb{N},
			\end{equation}
			where $R_{m,k}(r)$ is the $k$'th eigenfunction of the integral equation
			\begin{equation} \label{eq:integralEq}
			\lambda_{m,k}R_{m,k}(r) = \int_{0}^{a}h_m(r,r')R_{m,k}(r')dr',
			\end{equation}
			with $\{\lambda_{m,k}\}_k$ being the eigenvalues corresponding to $\{R_{m,k}\}_k$, and $h_m (r,r^{'})$ is given by
			\begin{equation*}
				h_m(r,r') = \int_{0}^{c}  G(\rho)J_m(\rho r')J_m(\rho r)r'\rho d\rho,
			\end{equation*}
			where $ J_m $ is the $m$'th order Bessel function of the first kind, $ c $ is the band-limit of $ f $, and $ G  $ is the RPSD of $ f $, given explicitly as
			\begin{equation}
			G(\rho) = \frac{1}{2\pi}	\int_{0}^{2\pi}\mathbb{E}|\hat{f}(\rho,\theta)|^2d\theta, \label{eq:G RPSD def}
			\end{equation}
			where $ \hat{f} = \mathcal{F}(f)$ is the two-dimensional Fourier transform of~$f$. It is worthwhile to point out that $\psi_{m,k}$ of~\eqref{eq:psi_hat steerable form} are \textit{steerable} functions~\citep{teo1999theory}, since correlating a template of the form of~\protect\eqref{eq:psi_hat steerable form} with an arbitrarily rotated image has absolute value which is independent of the rotation.
			
\begin{figure}
				\includegraphics[width=0.7\linewidth]{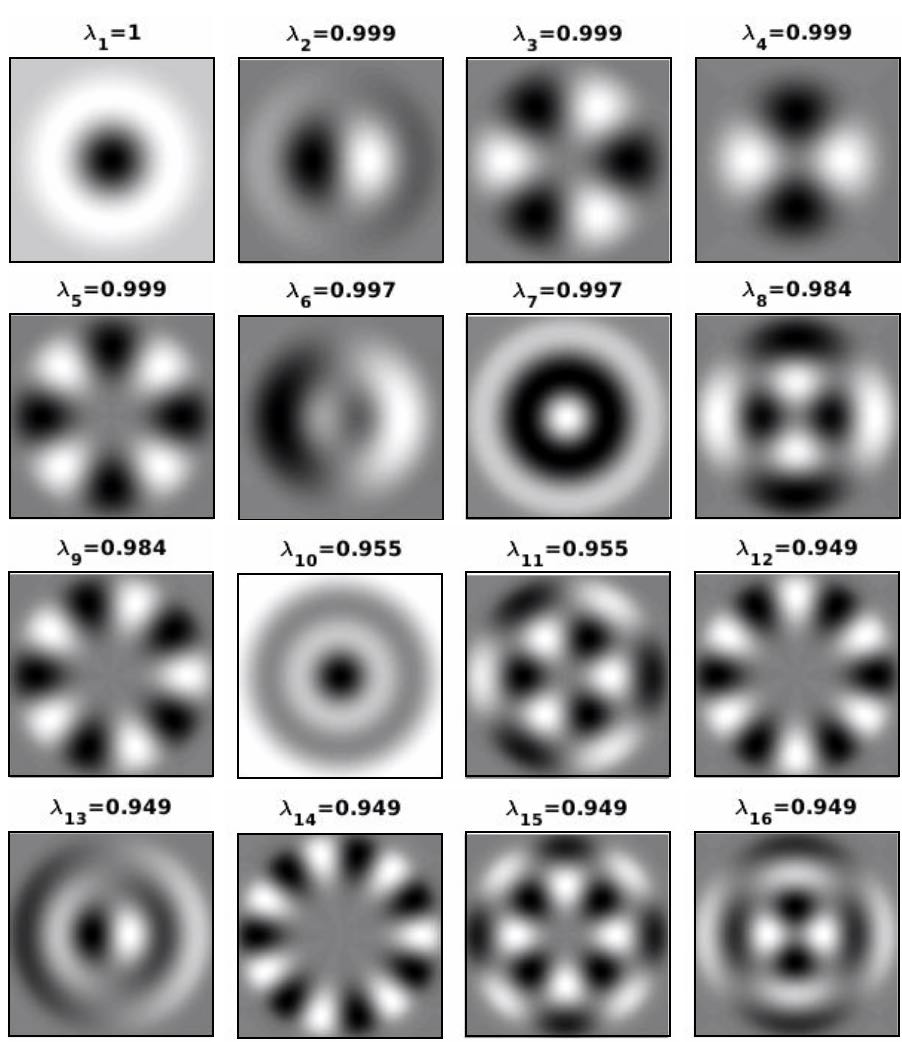}
				\centering
				\caption{The first 16 functions $\psi_n$ (for the largest $\lambda_n$) obtained from solving~\eqref{eq:integralEq} using the RPSD estimation procedure described in Section~\ref{subsec:psdEstimation}. The eigenfunctions and the eigenvalues were computed from a single typical micrograph of EMPIAR-10028 data set~\citep{PMID:24913268}, and were normalized so that the largest eigenvalue is one.} \label{fig:eigenfunctions}
			\end{figure}

\begin{figure}
				\includegraphics[width=0.9\linewidth]{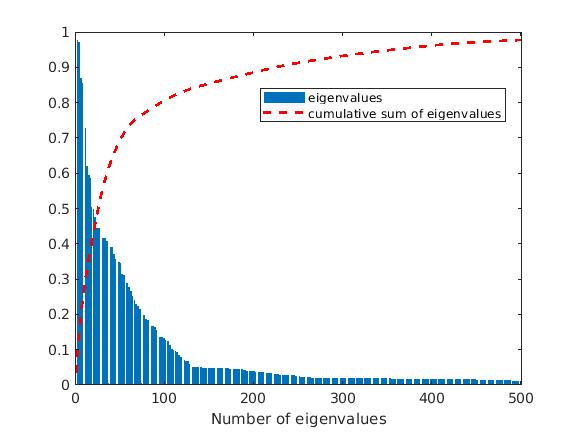}
				\centering
				\caption{The first 500 eigenvalues $\lambda_n$ (sorted in descending order) obtained from solving~\eqref{eq:integralEq} using the RPSD estimation procedure described in Section~\ref{subsec:psdEstimation}. The red dashed line is a normalized cumulative sum of the eigenvalues. The eigenvalues were computed from a single typical micrograph of EMPIAR-10028 data set~\citep{PMID:24913268} and were normalized so that the largest eigenvalue is one.
				It is evident that the eigenvalues decay rapidly, and $300$ eigenfunctions are already enough to capture more than $90\%$ of the variability in the particle images (since the normalized cumulative sum of the first 300 eigenvalues exceeds $0.9$).} \label{fig:eigenvalues}
			\end{figure}

			In order to solve~\eqref{eq:integralEq} and evaluate $R_{m,k}$, one needs to know the particle's RPSD $G $. In the following section we describe a method to estimate it for a given micrograph.
			Then, equipped with an estimate for $G $,  we use the well known Nyström's method \citep{atkinson1967numerical} in order to solve~\eqref{eq:integralEq} numerically. To simplify subsequent notation, we sort all eigenvalues $ \lambda_{m,k} $ (for all $ m $ and $ k $) in descending order, and denote the sorted eigenvalues by $ \lambda_n $. We sort the eigenfunctions accordingly. Figure~\ref{fig:eigenfunctions} illustrates the first 16  templates~$\psi_n$ (sorted according to their corresponding eigenvalues $\lambda_n$) obtained from the EMPIAR-10028 data set~\citep{PMID:24913268}, and Figure~\ref{fig:eigenvalues} depicts the first $500$ eigenvalues corresponding to the templates displayed in Figure~\ref{fig:eigenfunctions}.
						
		\subsection{Radial power spectral density estimation}\label{subsec:psdEstimation}
\begin{figure}
				\includegraphics[width=\linewidth]{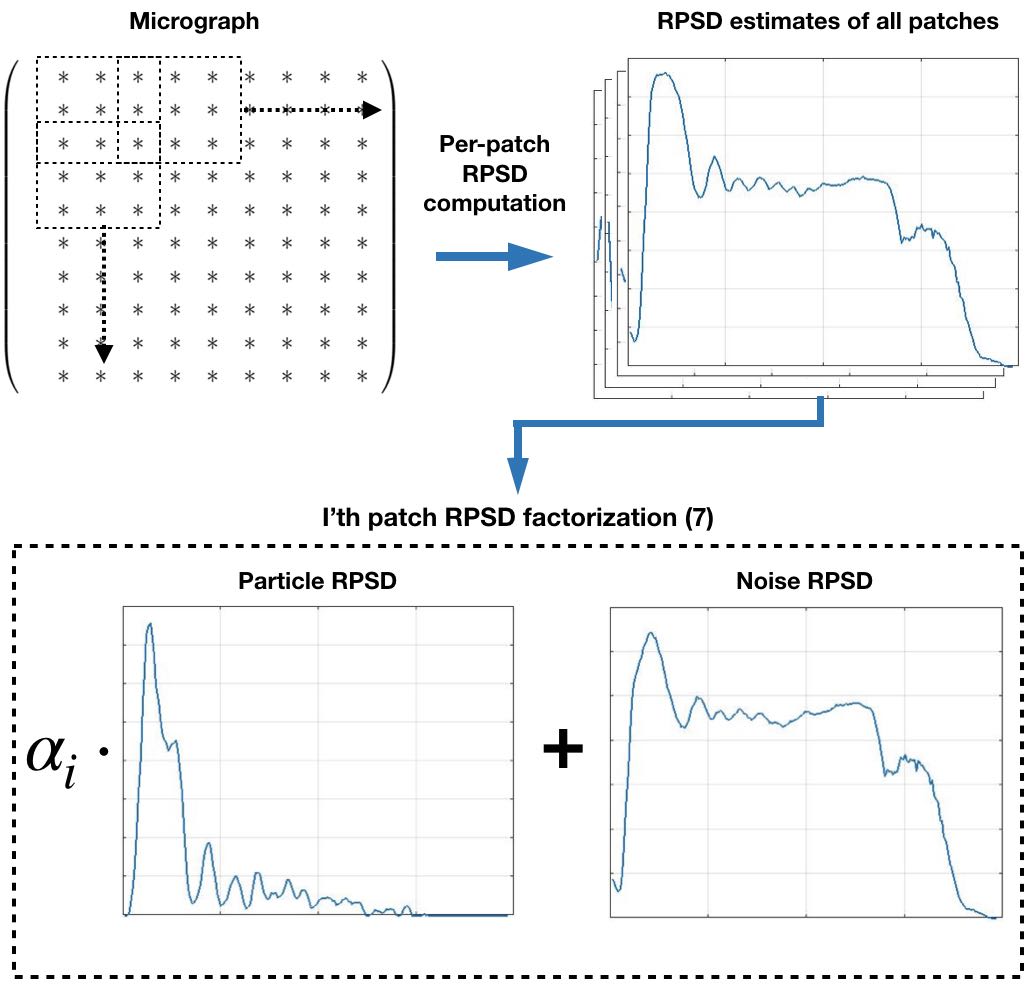}
				\centering
				\caption{Particle and noise RPSD estimation procedure. We divide the micrograph into (possibly overlapping)  $ l\times l $ patches where  $l$  is less than or equal to the maximal diameter of the particle. For each patch we compute its RPSD and factorize it as a sum of the noise RPSD and the particle RPSD multiplied by a proportion factor $ 0\leq\alpha\leq1$.}
				\label{fig:psdFactorization}
			\end{figure}

\begin{figure}
				\includegraphics[width=0.9\linewidth]{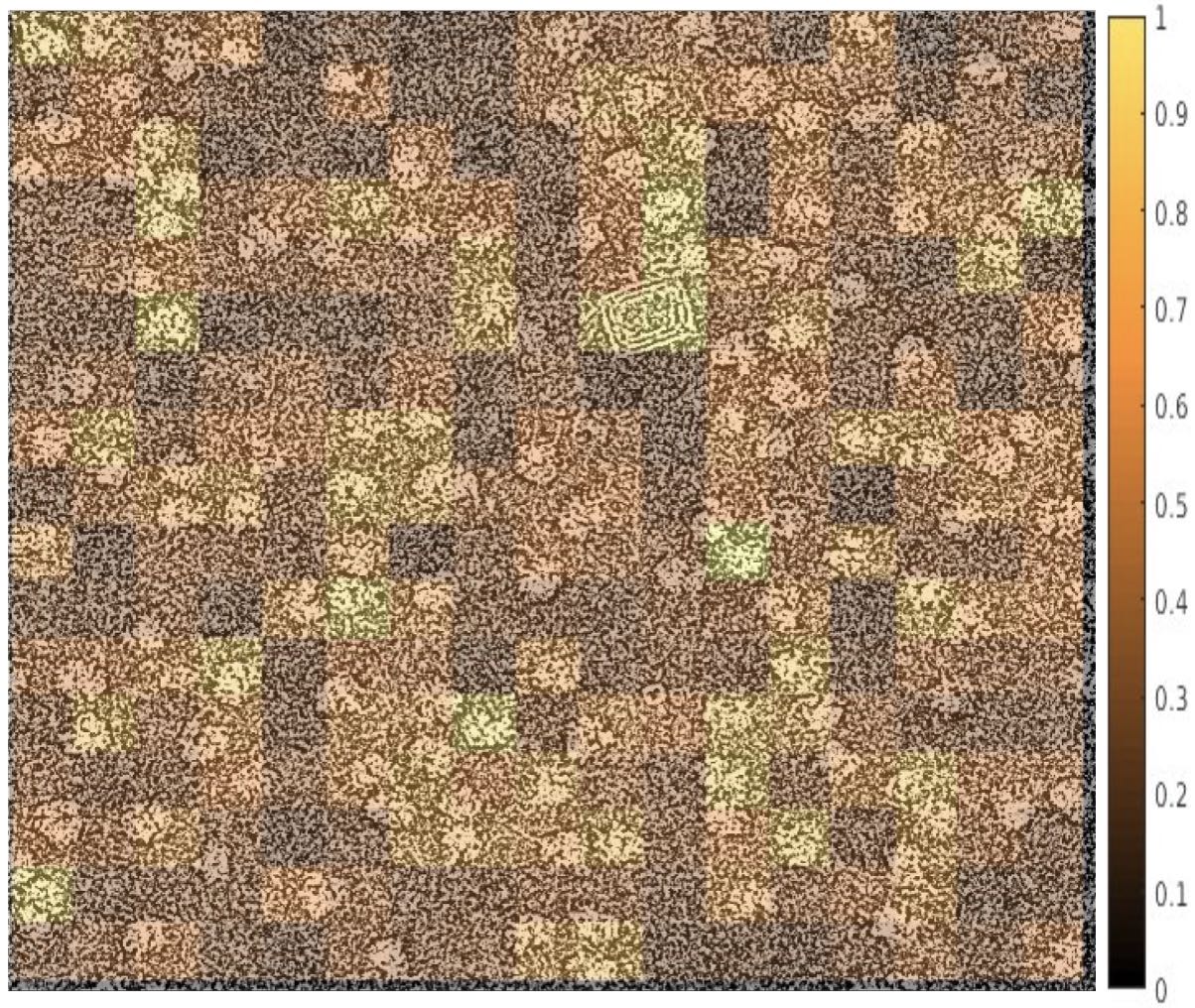}
				\centering
				\caption{Heat map of $ \{\hat{\alpha}_i \}$ over the micrograph it's estimated from. The values of $ \{\hat{\alpha}_i\} $ were obtained by solving~\eqref{eq:sFactor} for a typical micrograph of the EMPIAR-10028 data set~\citep{PMID:24913268}, and were normalized so that their largest value is~1. In order to create the heat map, $ \hat{\alpha}_i $ was used to color the $i$'th patch. As explained in Section~\ref{subsec:psdEstimation}, large values for $ \alpha_i $ (bright color) indicate the presence of a particle, and small values (dark color) indicate its absence.}
				\label{fig:alphaHeat}
		\end{figure}

			In this section, we describe a method for estimating the particle's RPSD $G $ (see~\eqref{eq:G RPSD def}) for a given micrograph.
			Towards this end, we partition the micrograph into $M$  (possibly overlapping) patches of size $l\times l$ with stride $d$. Specifically, if $A$  is a micrograph, then the first patch is defined by $ P_1 = A(1\colon l,1\colon l) $, the second by $ P_2=A(1\colon l,1+d\colon l+d) $ and so on (for a detailed explanation on how $ l,d $ are chosen see~\ref{app:estimating the RPSD}). Then, for each patch we estimate it's RPSD at $l$ equally-spaced points as explained in~\citep{papoulis1977signal}, and form the matrix $ \mathcal{S} \in \mathbb{R}_+^{l\times M} $ whose $ i' $th column $ \mathcal{S}_i $ is the RPSD computed from the $ i $'th patch.
			We consider the following model for the RPSD of the patches  in a given micrograph:
			\begin{align}\label{eq:psdPacthModel}
				\mathcal{S}_i = \nu +  \alpha_i \gamma,
			\end{align}
			where $\nu \in \mathbb{R}_+^l$ are the samples of the noise's RPSD, and  $\gamma \in \mathbb{R}_+^l $ are the samples of $G$ of~\eqref{eq:G RPSD def} (the particle's RPSD). We assume that $ \nu $ remains the same across all patches, and that $\gamma$ is multiplied by a factor $0\leq\alpha_i\leq1$ in each patch, which accounts for the proportion of the particle contained in the $i$'th patch. That is, a patch which contains no particle admits $\alpha=0$, a patch which fully contains a particle admits $\alpha=1$, and a patch which contains part of a particle admits $\alpha$ equal to the proportion of the particle in the patch.
			According to~\eqref{eq:psdPacthModel}, we can estimate $\nu$, $\gamma$ and $\{\alpha_i\}_{i=1}^M$ by solving
            \begin{equation}\label{eq:sFactor}
			\begin{gathered}
				\{\hat{\gamma},\hat{\nu},\hat{\alpha}\} = \underset{\tilde{\gamma},\tilde{\nu}, \tilde{\alpha}}{\operatorname{arg\,min}} \sum_{i=1}^M ||\mathcal{S}_i-\tilde{\nu} -\tilde{\alpha}_i \tilde{\gamma}||^2,\\
				\text{s.t.} \;\;\; \tilde{\gamma},\tilde{\nu}\in\mathbb{R}_+^l,\;\tilde{\alpha}_i\in[0,1],
			\end{gathered}
            \end{equation}
			where $ ||\cdot|| $ is the standard Euclidean norm.
			See Figure~\ref{fig:psdFactorization} for an illustration of the RPSD estimation process.
			Note that~\eqref{eq:sFactor} can be viewed as an instance of Non-Negative Matrix Factorization (NNMF)~\citep{gillis2017introduction} of the matrix $ \mathcal{S} $. In~\ref{app:estimating the RPSD}, we describe a method for estimating the solution of~\eqref{eq:sFactor} by using alternating least-squares \citep{kim2008nonnegative}, which is a popular approach for NNMF.
			We illustrate typical values of $\hat{\alpha}_i$ from the factorization~\eqref{eq:sFactor} in Figure~\ref{fig:alphaHeat}. It is evident that $\hat{\alpha}_i$ is larger in areas of the micrograph which are densely packed with particles, indicating that the factorization extracts the RPSD of the particles from the correct regions of the micrograph.
			
	 		\subsection{Particle detection}\label{subsec:particleDetection}
			In the previous sections, we described how to compute a set of templates $\{\psi_j\}$ (together with their eigenvalues $\{\lambda_j\}$ describing their importance) adapted to the particle images in the micrograph. However, at this point it is unclear how to make use of the different templates. In particular, since a single template does not capture the variability of all particle images, multiple templates must be used. We should therefore determine how many templates to use and how to combine them for particle picking. To that end, we turn to classical hypothesis testing and derive a scheme where different templates $\psi_j$ are combined according to their relative importance (through the eigenvalues $\lambda_j$).
			Specifically, we next derive a procedure for discriminating between an image of pure noise, which is defined as the null hypothesis and denoted by $H_0$, and an image of a particle plus noise, which is defined as the alternative hypothesis and denoted by $H_1$. That is, we consider the model
			 \begin{equation} \label{eq:statisticalModel}
	 			 \mathcal{I} =\begin{cases}
	 			 	\mathcal{N},       & H_0: \text{\space pure noise,}\\
					 \mathcal{P} + \mathcal{N}, &  H_1: \text{\space particle plus noise,}
				 \end{cases}
			 \end{equation}
			 where $ \mathcal{I} \in \mathbb{R}^{\lfloor2a+1\rfloor^2} $ is a flattened micrograph patch, $\mathcal{N} \in \mathbb{R}^{\lfloor2a+1\rfloor^2} $ is a flattened image of the noise only, $ \mathcal{P} \in \mathbb{R}^{\lfloor2a+1\rfloor^2}  $ is a flattened particle image without noise and with zero values outside of $ a\mathbb{D} $, and $ \lfloor\cdot\rfloor $ is the floor function.
		 	 We assume that the noise is white, which is enforced by prewhitening the micrograph using the samples of the estimated noise RPSD $\hat{\nu}$ (obtained in Section~\ref{subsec:psdEstimation}).
		 	 Given the model~\eqref{eq:statisticalModel}, we test for the hypotheses $ H_0 $ and $H_1$ by the well-known Likelihood Ratio Test (LRT)~\citep{wasserman2013all}.
			 We derive the LRT for the case where $ \mathcal{I}$ and $\mathcal{N} $ are normally distributed, and demonstrate the practical effectiveness of the resulting test in Section~\ref{sec:expRes}. To derive the LRT in this case, we need the covariance matrices for both hypotheses, namely $ \mathbb{E}\left[\mathcal{I}\mathcal{I}^T|H_0\right]$ and $ \mathbb{E}\left[\mathcal{I}\mathcal{I}^T|H_1\right] $.
			 Assuming that $ \mathcal{P}$ and $\mathcal{N} $ are independent, it follows that
			 \begin{equation}\label{eq:covMat}
			 \Sigma \equiv \mathbb{E}\left[\mathcal{I}\mathcal{I}^T|H_1\right] = \mathbb{E}\left[\mathcal{P}\mathcal{P}^T\right] + \sigma^2 I,
			 \end{equation}
			 where $\sigma$ is the standard deviation of the noise, and $ I $ is the ${\lfloor2a+1\rfloor^2} \times {\lfloor2a+1\rfloor^2} $ identity matrix. Since $\sigma$ is typically unknown, we propose a method for estimating it from a given micrograph in~\ref{app:noise varience estimation}.	
			To evaluate~\eqref{eq:covMat} it therefore remains to estimate $\mathbb{E}\left[\mathcal{P}\mathcal{P}^T\right]$.
			Let $ U $ be an equally-spaced (Cartesian) grid with $ \lfloor2a+1\rfloor \times \lfloor2a+1\rfloor $ points over the square $ \lfloor-a,a\rfloor\times \lfloor-a,a\rfloor $, and denote by $ \mathcal{U}\in\mathbb{R}^{\lfloor2a+1\rfloor^2} $ the flattened version of $U$. By setting the values of $ f $ outside of $ a\mathbb{D} $ to be zero,  we have that $ \mathcal{P}_{i} = f(i) $ for every $i\in \mathcal{U}$. Setting also the values of $\psi_j$ of~\eqref{eq:templatesDef} outside of $ a\mathbb{D} $ to zero, and denoting by $\Psi\in\mathbb{R}^{\lfloor2a+1\rfloor^2 \times N}$ the matrix whose $j$'th column consists of the flattened samples of $\psi_j$  on $\mathcal{U}$, leads to the following approximation  of the particle's covariance matrix (see~\ref{app:LRT derivation})
			\begin{equation}\label{eq:cleanCovApprox}
				\mathbb{E}\left[\mathcal{P}\mathcal{P}^T\right] \approx \Psi\Lambda\Psi^T,
			\end{equation}
			where $ \Lambda $ is a diagonal matrix with the eigenvalues $ \{\lambda_j\}_{j=1}^N $ on its main diagonal. The parameter $ N  $ is chosen such that the error in the approximation~\eqref{eq:cleanCovApprox} is small, which essentially depends on the decay rate of the eigenvalues $\lambda_j$ (see~\ref{app:LRT derivation} for a detailed explanation on how the parameter $ N $ is determined).
			
\begin{figure}
			\centering
 			\includegraphics[width=\linewidth]{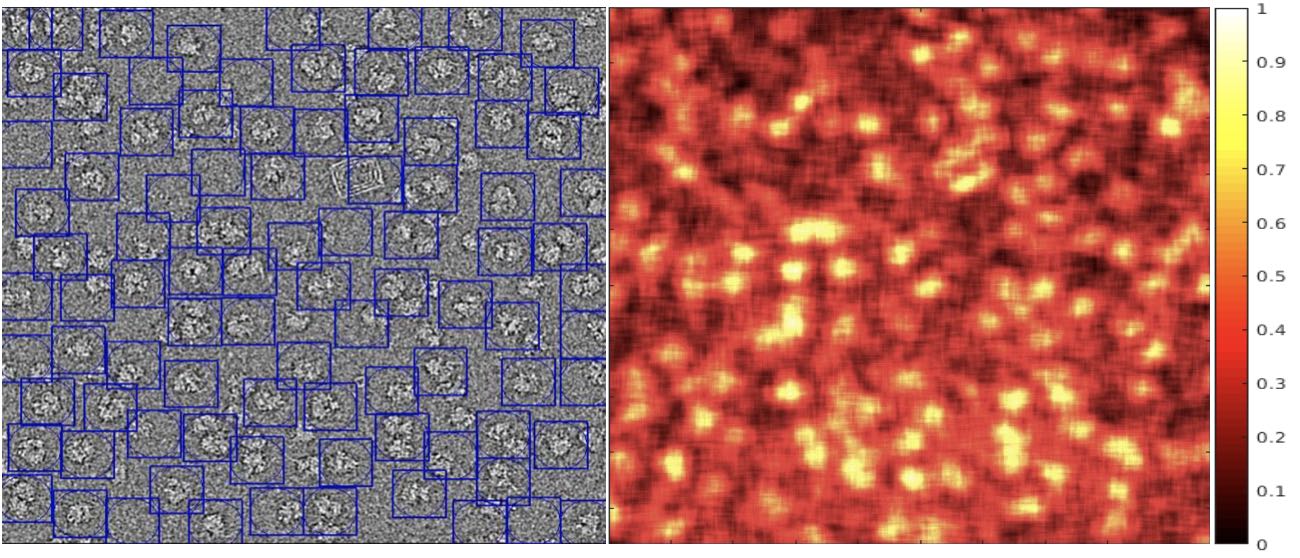}
			\caption{Example of a typical scoring matrix as a heat map (right image) and particles picked from it (left image) with a threshold of zero. The data set is EMPIAR-10028~\citep{PMID:24913268}.}
			\label{fig:scoringMat}
		\end{figure}

\begin{figure*}
		\includegraphics[width=0.8\textwidth]{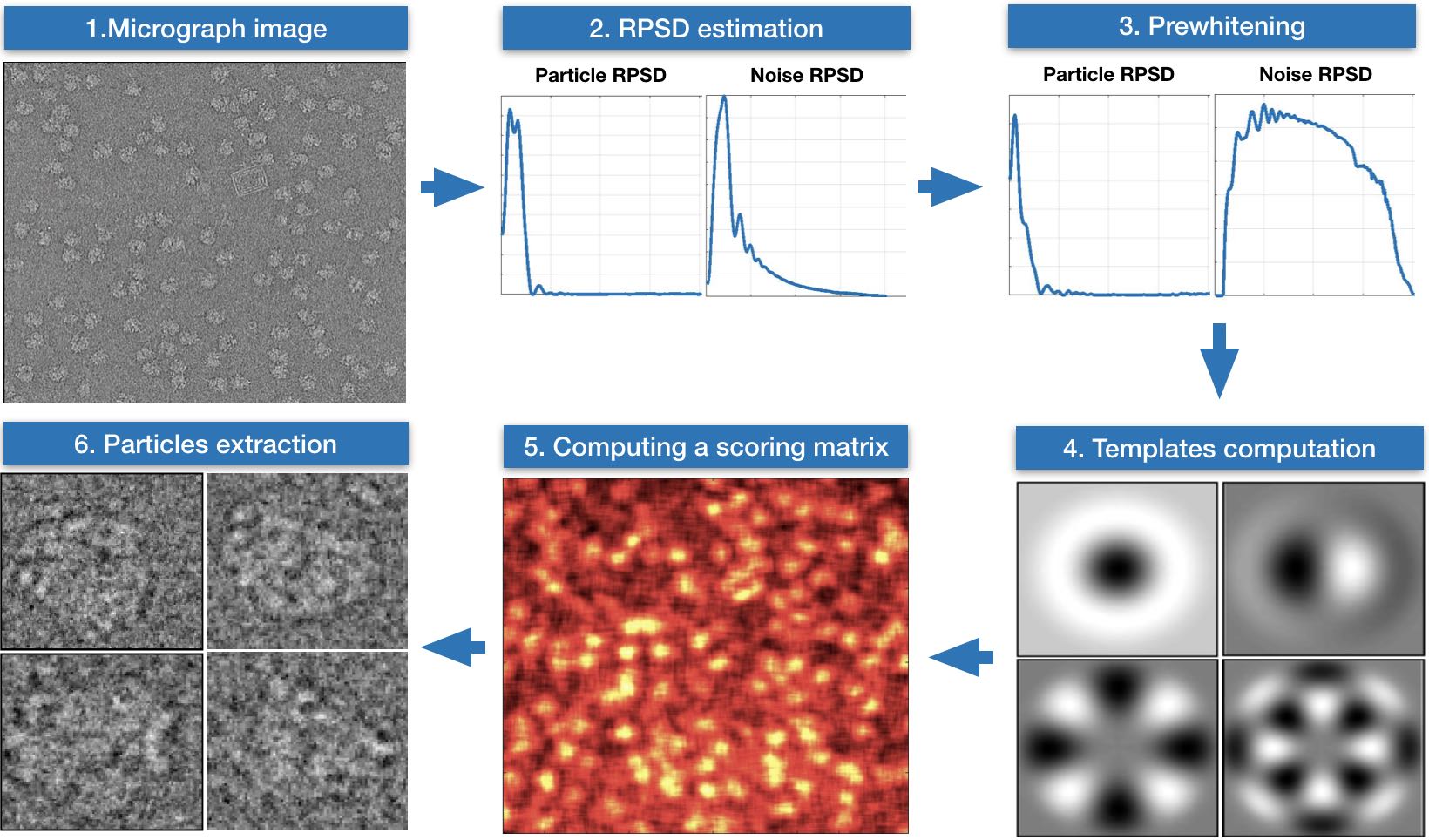}
		\centering
		\captionsetup{width=1\linewidth}
		\caption{For each micrograph, we estimate the particle and noise  RPSD as explained in Section~\ref{subsec:psdEstimation}. Then, we prewhiten the micrograph, and estimate the templates using~\eqref{eq:integralEq}. Last, we compute the LRT scoring matrix and extract the particles' coordinates from it, as described at the end of Section~\ref{subsec:particleDetection}.}
		\label{fig:diagramOfAlgo}
	\end{figure*}

			Given the covariance matrix $\Sigma$ of~\eqref{eq:covMat} (computed via the approximation~\eqref{eq:cleanCovApprox}), we use a score based on the LRT for comparing between two multivariate normal distributions as follows. Each micrograph patch $ \mathcal{I}^{i,j} $ of size $ \lfloor2a+1\rfloor\times \lfloor2a+1\rfloor $ (flattened to a vector) is considered as a realization of $ \mathcal{I} $ and is assigned a score
			\begin{align}\label{eq:scoringMatrix}
				\text{score}(i,j)&:=\log\left(\frac{p(\mathcal{I}^{i,j}|H_1)}{p(\mathcal{I}^{i,j}|H_0)}\right)\\
				&=\frac{1}{2}\left(\log\left(\frac{|\Sigma|}{\sigma^{2\lfloor2a+1\rfloor^2}}\right)+\left(\mathcal{I}^{i,j}\right)^T\left(\sigma^{-2}I-\Sigma^{-1}\right)\mathcal{I}^{i,j}\right), \nonumber
			\end{align}
where  $ |\cdot| $ is a matrix determinant. See~\ref{app:LRT derivation} for an elaborate derivation of this score and an efficient algorithm to compute it.
High value of 	$\text{score}(i,j)$ indicates the presence of a particle in the $(i,j)$ patch. We consider the scores as a matrix whose elements are $\text{score}(i,j)$.
Extracting the particle's coordinates out of the scoring matrix is done by using a version of the non-maximum suppression algorithm~\citep{neubeck2006efficient}.
			This greedy algorithm first chooses the highest score in the scoring matrix as the first picked particle's center. Then, in order to prevent nearby pixels from being considered as particles, all pixels within a  user-defined radius are excluded from the picking, and the procedure resumes with the next largest score, and so on. We set the exclusion radius to be the particle size, however, smaller values may give better results for closely packed, irregularly shaped particles.
			The picking continues until the remaining scores are below zero or until the algorithm picks a predefined number of particles. Stopping when the remaining scores are negative is based on the observation that $\text{score}(i,j)<0$ holds whenever $p(\mathcal{I}^{i,j}|H_1)<p(\mathcal{I}^{i,j}|H_0)$, which means that all of the remaining patches have higher probability to contain noise only.
			See Figure~\ref{fig:scoringMat} for an example of a scoring matrix and the particles picked from it.
		See Figure~\ref{fig:diagramOfAlgo} and algorithm~\ref{alg:particlePicking} for a diagram and an algorithm that summarize the particle picking process.
	
	\begin{algorithm}
		\caption{\textbf{The KLT picker}}
		\label{alg:particlePicking}
		\begin{algorithmic} [1]
			\Statex{\textbf{Required:} Micrograph and the particles' radius $ a $.}
				\State Partition the micrograph into $ l\times l $ patches with stride $ d $ in each dimension (by default, $ l $ is chosen to be 80\% of the maximal diameter of the particle and $ d=l $, see~\ref{app:estimating the RPSD} for more details).
            \State Compute the RPSD of each patch as described in section~\ref{subsec:psdEstimation}.
			\State Estimate the particle and noise RPSD $ \hat{\gamma},  \hat{\nu} $, respectively by solving~\eqref{eq:sFactor}.
			\State  Prewhiten the micrograph using $ \hat{\nu} $.
				\State Compute the set of eigenfunctions $ \{\hat{\psi_j}\} $ and eigenvalues $ \{\lambda_j\} $ of~\eqref{eq:templatesDef} by solving~\eqref{eq:integralEq} using $ \hat{\gamma} $ and the Nystr\"{o}m method.
			\State Estimate the particle's covariance matrix $ \mathbb{E}[\mathcal{P}\mathcal{P}^T]$  according to~\eqref{eq:cleanCovApprox} and the noise standard deviation $ \sigma $, as detailed in~\ref{app:noise varience estimation}.
			\State Compute $ \Sigma = \mathbb{E} \left[\mathcal{I}\mathcal{I}^T|H_1\right] $ (see~\eqref{eq:covMat}).
			\State Partition the micrograph into $ [2a+1]\times [2a+1] $ patches with stride $ 1 $ in each dimension and compute the LRT score for each patch according to~\eqref{eq:scoringMatrix}. The outcome is a scoring matrix.
			\State Extract the particles' coordinates from the scoring matrix as described at the end of Section~\ref{subsec:particleDetection}.
		\end{algorithmic}
	\end{algorithm}

	\section{Experimental results}\label{sec:expRes}
	In this section, we demonstrate the performance of our algorithm on three different data sets, as follows. First, the micrographs were downsampled to 20\%-50\% of their original size and were normalized to zero mean and variance of one. In addition, each micrograph was bandpass filtered, eliminating 5\%  of its lowest and highest radial frequencies. The lowest frequencies were eliminated since they do not contain enough samples for estimating the RPSD, and the highest frequencies were eliminated since they contain mostly noise. Throughout all of the experiments, we used the following parameters for the KLT picker: $a$ and $l$ are 20\% and 80\%, respectively, of the maximal diameter of the particle, $ d=l $, and as explained above, the detection threshold is set to zero. Note that $ a $ is taken to be smaller than half of the estimated particle's size, as this leads, in practice, to better centered particle selections. The obtained particle coordinates were imported into RELION~\citep{scheres2015semi} and the routine described in~\citep{scheres2014single} was executed in order to obtain a 3D reconstruction. The EMDB~\citep{EMDB} model corresponding to each data set was used as the ab initio model for the 3D refinement process.
	
	The procedure described above is demonstrated on the following data sets of the EMPIAR repository~\citep{EMPIAR}: EMPIAR-10028 (Plasmodium Falciparum 80S ribosome) ~\citep{PMID:24913268}, EMPIAR-10061 ($ \beta $-Galactosidase)~\citep{PMID:25953817}, and EMPIAR-10049 (Synaptic RAG1-RAG2)~\citep{PMID:26548953}.
	
	Our algorithm was implemented in Matlab, and all the experiments were executed on a Linux machine with 16 cores running at 2.1GHz, 768GB of memory, and an Nvidia Titan GPU.
	
	\subsection{Plasmodium Falciparum 80S}\label{subsec:Plasmodium}
\begin{figure*}
			\includegraphics[width=1\textwidth]{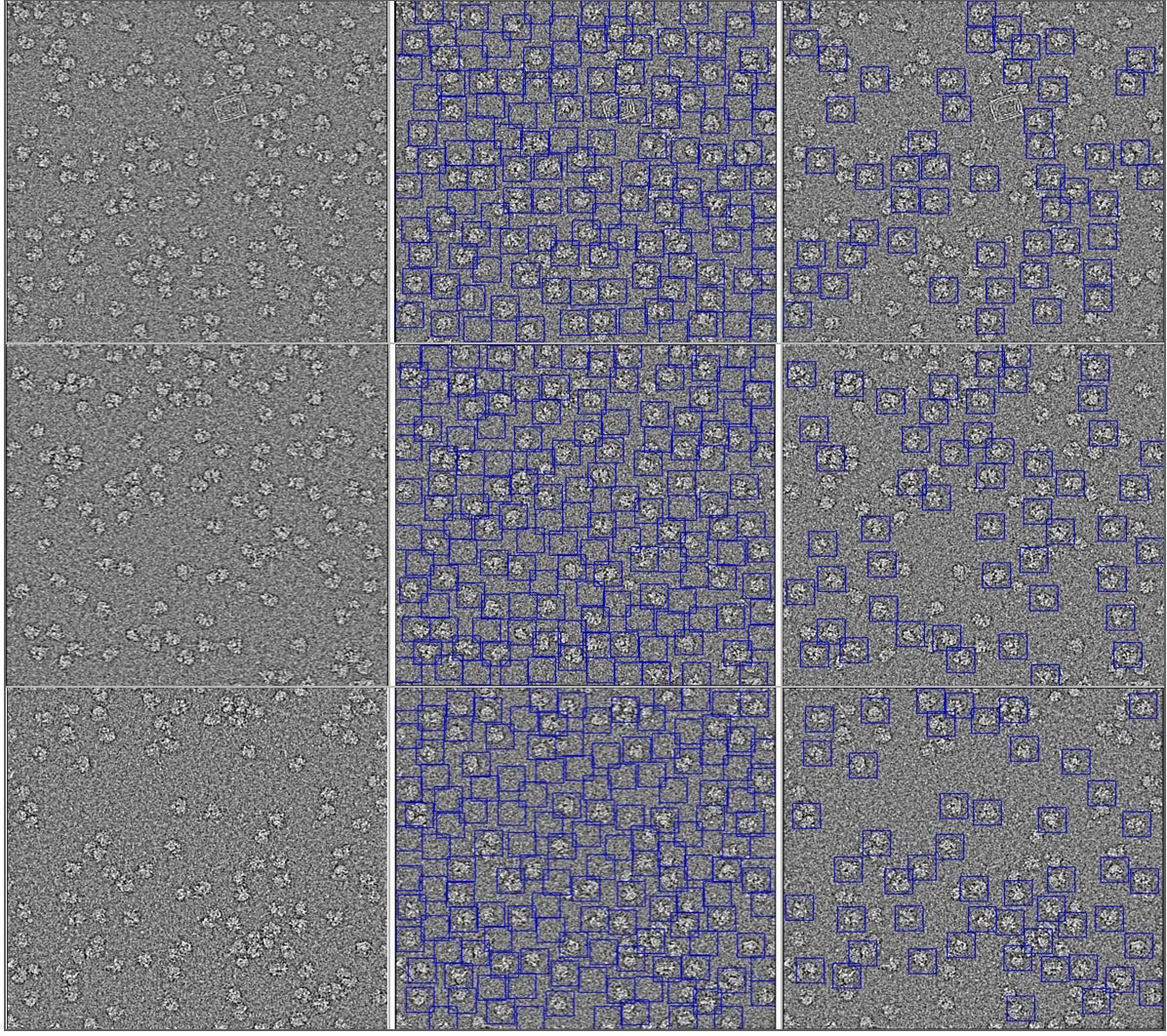}
			\centering
			\captionsetup{width=1\linewidth}
			\caption{Examples of picking results of the Plasmodium Falciparum 80S data set. The micrographs are shown in the left column, picked particles with threshold equal to zero are in the middle columns, and the top 50 picked particles are in the right column.}
			\label{fig:10028PickedPar}
		\end{figure*}

		The Plasmodium Falciparum 80S data set consists of $1081$ micrographs of size  $4096\times 4096$ pixels, with pixel size of $1.34~\AA$. The approximate size of the particle is $360 \times 360$ pixels (according to~\citep{PMID:24913268}).
		Examples of the KLT picking results are shown in Figure~\ref{fig:10028PickedPar}.
		The KLT picker picked  $141,679$ particles from the micrographs.  In order to purge the set of picked particles, a 2D class averaging step was executed  (using $30$ classes) and $87,439$ particles retained. Examples of the selected 2D classes are shown in Figure~\ref{fig:class2D}. The 3D reconstruction reached a gold standard Fourier Shell Correlation (FSC) resolution of $3.05 {\AA} $. Figure~\ref{fig:fscCurve} presents the FSC curve produced by RELION's post-processing task. In comparison, $158,212$ particles were picked by~\citep{PMID:24913268} using RELION's automated selection tool, and after 2D and 3D class averaging steps, $105,247$  particles remained. The reconstruction in~\citep{PMID:24913268} reached a gold standard FSC resolution of $3.20 {\AA} $.
		Figure~\ref{fig:surface} presents surface views of our 3D reconstructed model and the reconstructed model of ~\citep{PMID:24913268}.
		The running time of the KLT picker for a single micrograph, using a GPU, was approximately $ 35 $ seconds (following a one-time preprocessing step of approximately $ 40 $ seconds).
		\subsection{$\beta$-Galactosidase}\label{betaGalactosidase}
\begin{figure*}
			\includegraphics[width=1\textwidth]{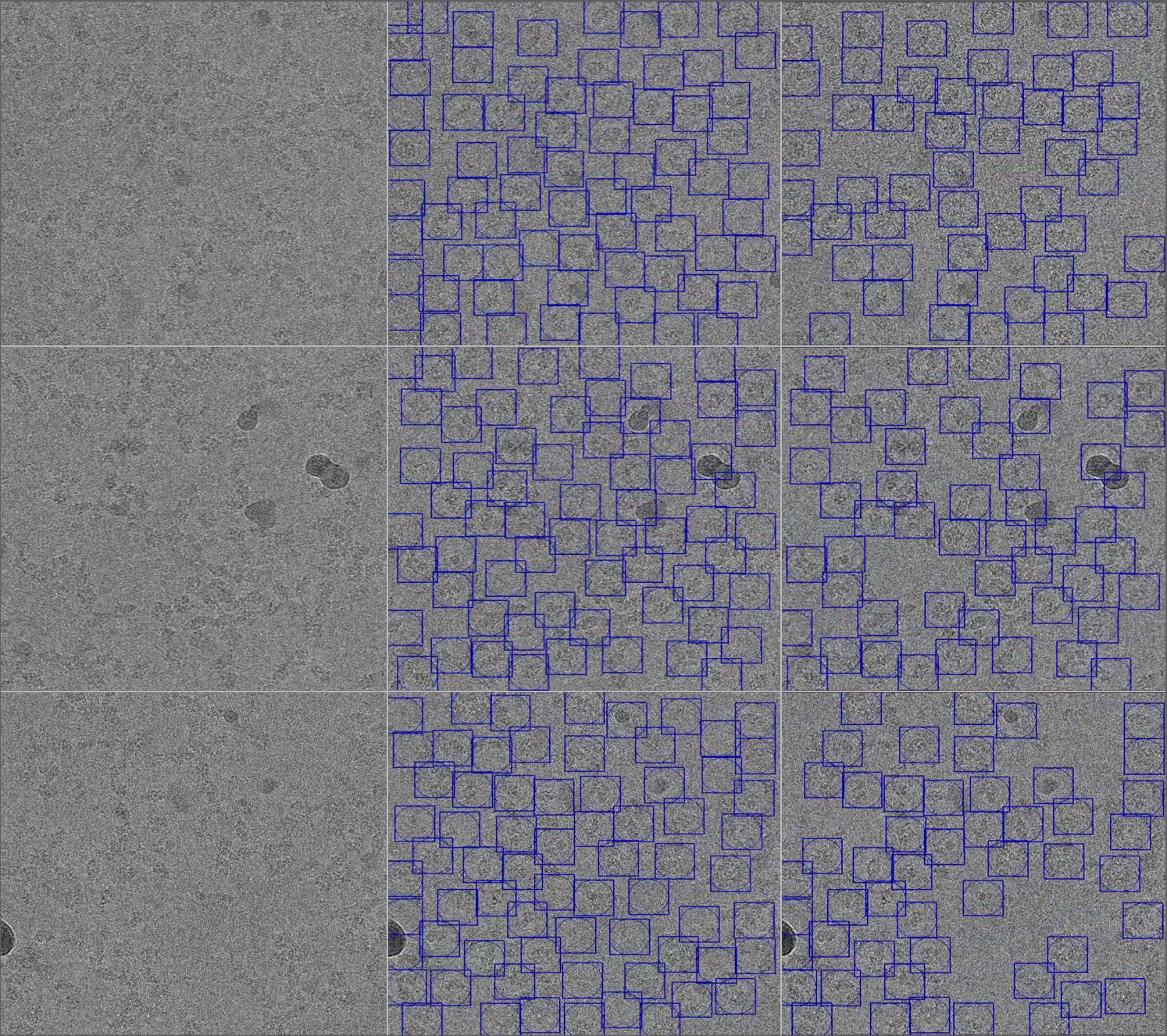}
			\centering
			\captionsetup{width=1\linewidth}
			\caption{Examples of picking results of the $\beta$-Galactosidase data set. See Figure~\ref{fig:10028PickedPar} for details.}
			\label{fig:10061PickedPar}
		\end{figure*}

		The $ \beta $-Galactosidase data set consists of 1539 micrographs, each of size $ 7676 \times 7420 $ pixels, with pixel size of 0.64$ \AA $. The approximate size of the particle according to~\citep{PMID:25953817} is $ 768 \times 768$ pixels. KLT picker picked $100,241$ particles. Examples of the results of the picking are shown in Figure~\ref{fig:10061PickedPar}. After executing a 2D class averaging step (using $30$ classes), $13,000$ particles retained. Examples of the selected 2D classes are shown in Figure~\ref{fig:class2D}. The 3D reconstruction reached a gold standard FSC resolution of $3.01 {\AA} $. Figure~\ref{fig:fscCurve} presents the FSC curve produced by RELION's post-processing task. It is important to note that the micrographs' raw movies were not available to us, therefore better results may be achieved by further polishing. In comparison,~\citep{PMID:25953817} picked $ 93,686 $ particles using an automatic method based on correlation with a Gaussian ring. Than, a 3D classification in FREALIGN~\citep{grigorieff2007frealign} was used to discard bad selections, and $ 41,123 $ particles retained for the 3D refinement step (that was executed in the same program). The reconstruction in~\citep{PMID:25953817} reached a gold standard FSC resolution of $2.2 {\AA} $. Figure~\ref{fig:surface} presents surface views of the 3D models.
		The running time for a single micrograph, using a GPU, was approximately $33$ seconds (following a one-time preprocessing step of approximately $40$ seconds).
 		
		\subsection{Synaptic RAG1-RAG2}\label{subsec:Synaptic}
\begin{figure*}
			\includegraphics[width=1\textwidth]{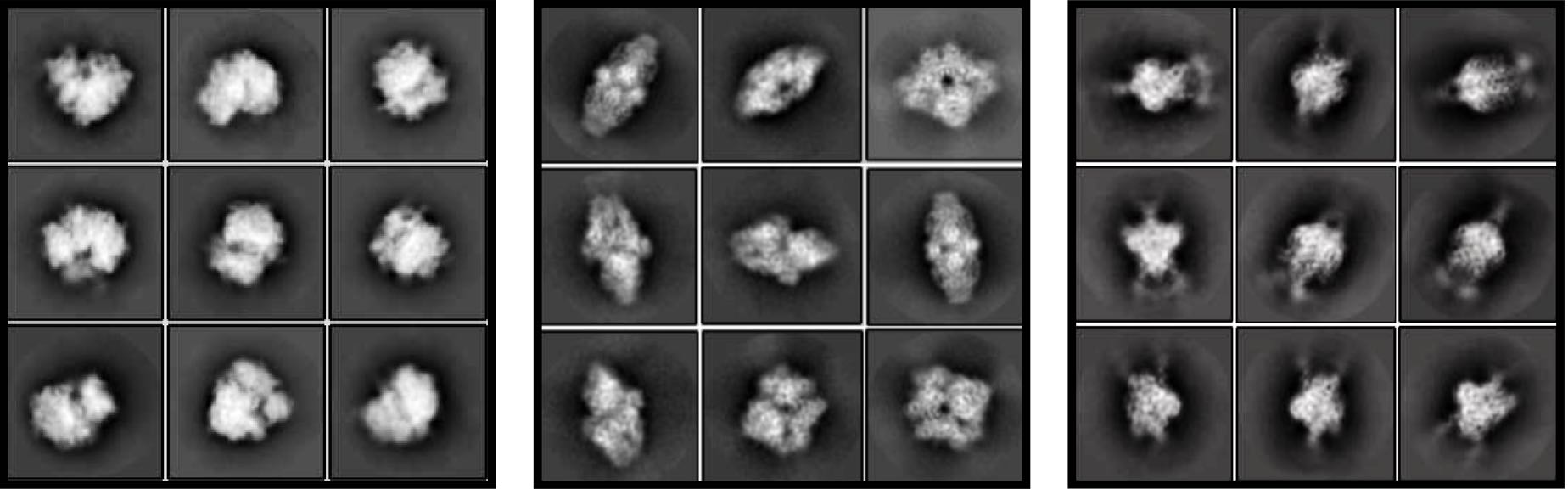}
			\centering
			\captionsetup{width=1\linewidth}
			\caption{Examples of $9$ 2D class averages of  Plasmodium Falciparum 80S (left),  $ \beta $-Galactosidase(middle), and  Synaptic RAG1-RAG2 (right) produced by RELION from particles picked by the KLT picker.}
			\label{fig:class2D}
		\end{figure*}

\begin{figure*}
			\includegraphics[width=1\textwidth]{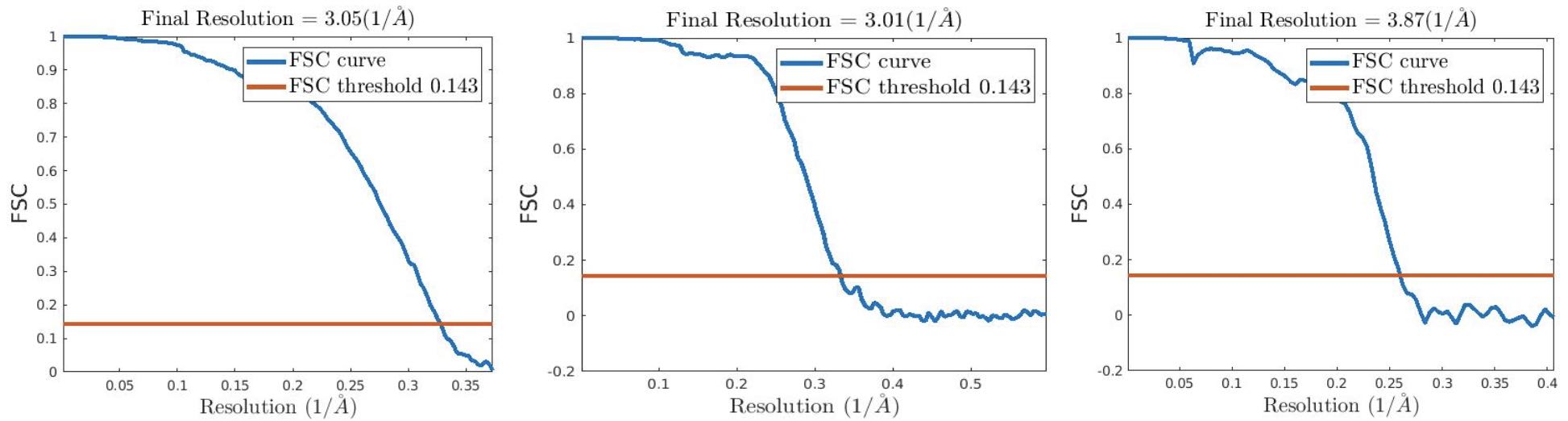}
			\centering
			\captionsetup{width=1\linewidth}
			\caption{FSC curves of Plasmodium Falciparum 80S (left), $ \beta $-Galactosidase  (middle), and Synaptic RAG1-RAG2 (right) produced by RELION's post-processing task.}
			\label{fig:fscCurve}
		\end{figure*}

\begin{figure*}
			\includegraphics[width=1\textwidth]{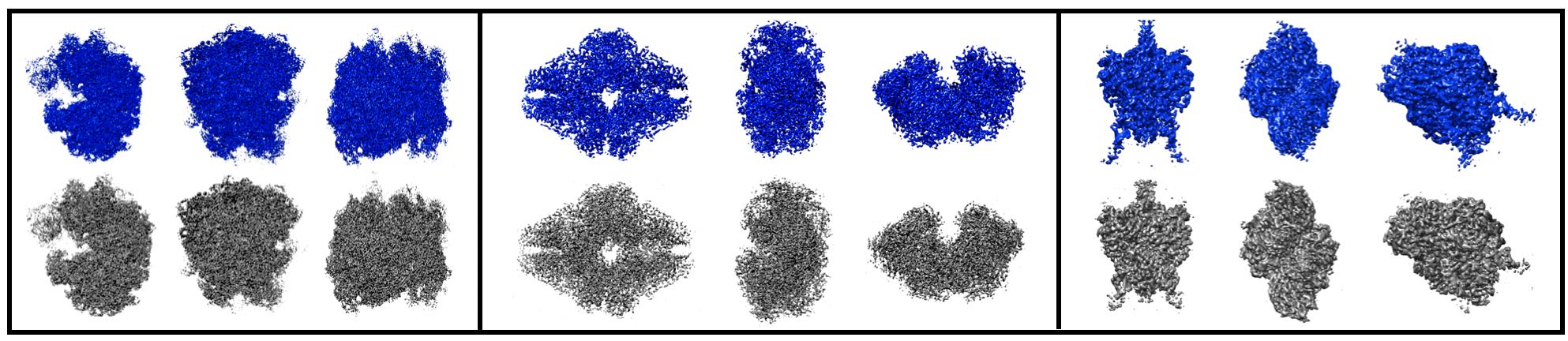}
			\centering
			\captionsetup{width=1\linewidth}
			\caption{Surface views taken from the 3D reconstructions of Plasmodium Falciparum 80S (left), $ \beta $-Galactosidase (middle), and Synaptic RAG1-RAG2 (right) . The top surface views correspond to reconstructions from particles picked by the KLT picker, and the bottom ones to the reconstruction of~\citep{PMID:24913268},~\citep{PMID:25953817}, and~\citep{PMID:26548953}, respectively. The views were generated using UCSF Chimera software~\citep{pettersen2004ucsf}.}
			\label{fig:surface}
		\end{figure*}

\begin{figure*}
			\includegraphics[width=1\textwidth]{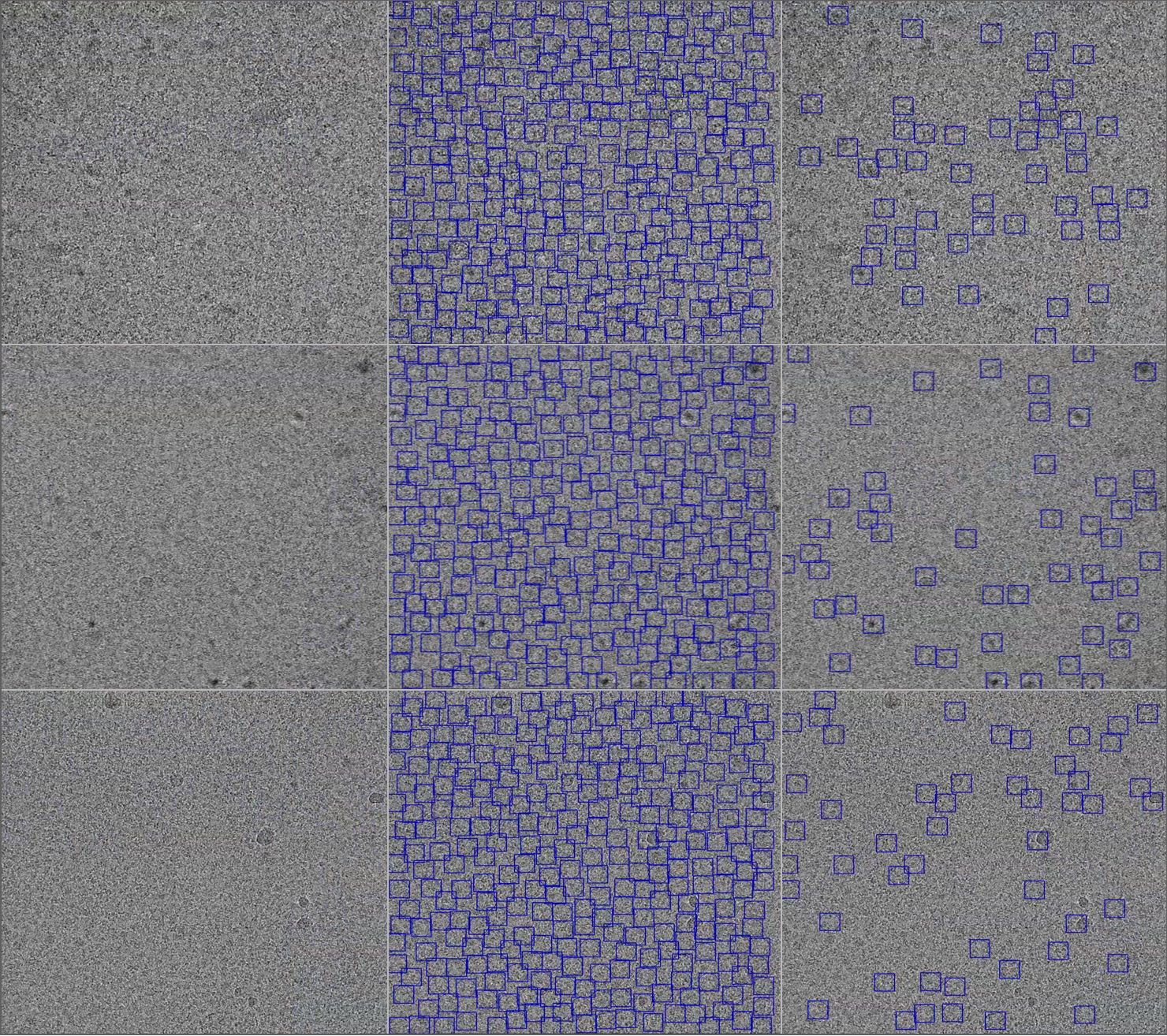}
			\centering
			\captionsetup{width=1\linewidth}
			\caption{Examples of picking results of the Synaptic RAG1-RAG2 data set.  See Figure~\ref{fig:10028PickedPar} for details.}
			\label{fig:10049PickedPar}
		\end{figure*}

		The Synaptic RAG1-RAG2 consists of 680 micrographs, each of size $ 3838 \times 3710 $ pixels, with pixel size of 1.23$ \AA $. The approximate size of the particle according to~\citep{PMID:26548953} is $ 192 \times 192$ pixels.
		The KLT picker picked $163,470$ particles. Examples of the results of the picking are shown in Figure~\ref{fig:10049PickedPar}. After executing a 2D class averaging step (using $30$ classes), $34,100$ particles retained. Examples of the selected 2D classes are shown in Figure~\ref{fig:class2D}. The 3D reconstruction reached a gold standard FSC resolution of $3.87 {\AA} $. Figure~\ref{fig:fscCurve} presents the FSC curve produced by RELION's post-processing task. We note that similarly to the $ \beta $-Galactosidase data set, the raw movies were unavailable to us, hence the results are sub-optimal. In comparison,~\citep{PMID:26548953} used a semi-automatic method executed in SAMUEL~\citep{shi2008script} for the particle picking step. Initially, $ 2,000 $ particles were manually selected and classified by a 2D class averaging step. Good classes were manually chosen and the particles in those classes were used as templates for an automatic picking process. As described in~\citep{PMID:26548953}, additional  manual effort was required to discard bad selections, which resulted in a final data set of  $89,158$ particles. The 3D refinement step was executed in RELION, and the reconstruction reached a gold standard FSC resolution of $3.4 {\AA} $. Figure~\ref{fig:surface} presents surface views of the 3D model.
		The running time  for a single micrograph, using a GPU, was approximately $50$ seconds (following a one-time preprocessing step of approximately $40$ seconds).
	

	\section{Discussion}
		\subsection{Review of the results}
		The  data sets used to demonstrate our method represent different types of micrographs. The Plasmodium Falciparum 80S  is the least challenging, as the macro-molecule is large ($ 4.2 MDa $) and the particles can easily be spotted by a naked eye. The $ \beta $-Galactosidase is more challenging, as the macro-molecule is smaller ($ 0.465 MDa $), and it is harder to see the particles on the micrographs. Finally, the Synaptic RAG1-RAG2 is the most challenging data set, as it corresponds to smaller macro-molecule ($ 0.38 MDa $) and has low SNR.
		With all these data set, we achieved high resolutions with minimum manual effort. In addition, the surface views in Figure~\ref{fig:surface} show that the 3D reconstructions we got are compatible with the published ones.
		
		Throughout the entire reconstruction process, the only manual step was the selection of classes out of RELION's 2D classification output. We did not use 3D classification, nor any other method to filter the KLT picker's picked particles. It is also important to note that the parameters in all of the experiments where chosen automatically based on the particle's size, so there is no need to adjust them manually.
		\subsection{Comparison with other particle picking methods}
		Particle picking methods can be divided into two categories. The first relies on prior assumptions on the particle's structure, such as Gaussian shape~\citep{voss2009dog} or resemblance to other known macro-molecules~\citep{wang2016deeppicker,xiao2017fast}. The second relies solely on information from the data itself, which can be a set of manually picked particles~\citep{scheres2015semi,bepler2018positive}, or a set of micrograph patches that were automatically selected~\citep{heimowitz2018apple}.
		The KLT picker is in between both categories. On one hand, it relies on the assumption that the projection images can be approximated by a band-limited, rotationally invariant, wide sense stationary process restricted to a disk. This assumption is non-restrictive in practice, as exemplified by our results. On the other hand, the templates of the KLT picker are adapted to every micrograph separately, through the RPSD of the particles and of the noise.

	\section{Conclusion}
	In this paper, we presented the KLT picker, an automated, data driven particle picker, that requires only one input which is the estimated particle's size. The KLT picker is designed especially to handle low SNR micrographs, it is  easy to use, fast, and mathematically rigorous. We evaluated the performance of the KLT picker on three different data sets. In all of them we achieved high-quality results with minimal manual effort.
	
\begin{figure}
		\includegraphics[width=0.8\linewidth]{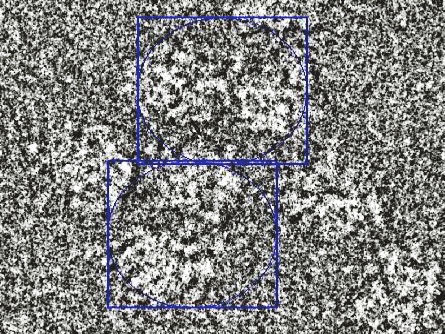}
		\centering
		\caption{An example of a dense particles cluster for which the KLT picker picked the interface between particles instead of the center of each particle. Micrograph taken from EMPIAR-10028 data set~\citep{PMID:24913268}.} \label{fig:densePar}
	\end{figure}

	The algorithm can be further improved as follows.
	Micrographs tend to contain contaminations of various forms, from small contaminations to whole areas containing either ice or the edges of the supporting grid. These areas can degrade the estimation of the RPSD, which is the essence of our method.  Devising an algorithm to automatically identify these areas may improve the picking.
	Dense particle clusters can poses a problem to the KLT picker, as the scoring matrix, introduced in Section~\ref{subsec:particleDetection}, tends to assign high scores to the interface between particles instead of to the center of each particle separately, see an example in Figure~\ref{fig:densePar}.
	One possible solution to this problem is to replace the current algorithm for extracting particles, which uses the individual values of the scoring matrix (described at the end of Section~\ref{subsec:particleDetection}), with an algorithm that considers the entire neighborhood of a particle candidate.

	\section*{Acknowledgments}
This research was supported by the European Research Council (ERC) under the European Union's Horizon 2020 research and innovation programme (grant
agreement 723991 - CRYOMATH).

	\appendix
	\section{Deriving~\eqref{eq:psi_hat steerable form} and~\eqref{eq:integralEq}}\label{app:templates derivation}
	The Karhunen Loeve Theorem~\citep{maccone2009simple} states that the solutions $ \psi_n$ to~\eqref{eq:templatesDef} are the eigenfunctions of the autocorrelation operator $ K(x,y) = \mathbb{E}\left[f(x)f(y)\right] $, that is,
	\begin{equation} \label{app:eigenFunEq}
		\lambda_n\psi_n(x)=\int_{a\mathbb{D}}K(x,y)\psi_n(y)dy.
	\end{equation}
	Under the assumptions that $ f $ is band-limited, wide sense stationary, and that $ K $ is radially-symmetric \big(i.e $ K(x,y) = K(\norm{x-y}) $\big), the Wiener–Khinchin theorem~\citep{cohen1998generalization} states that the radial power spectral density (RPSD) of $ f $, given by~\eqref{eq:G RPSD def}, is the Fourier transform of the autocorrelation $K$. Hence, by taking the inverse Fourier transform,
	\begin{equation}\label{app:FourierOfK}
		K(\norm{x-y}) = \frac{1}{(2\pi)^2}\int_{c\mathbb{D}}G(\norm{\omega})e^{i\omega\cdot(x-y)}d\omega,
	\end{equation}
	 where $ c $ is the band-limit of $ f $. Denote $x = (r\cos\theta ,r\sin\theta)$, $ y =(r'\cos\theta' ,r'\sin\theta') $, $  \omega =(\rho \cos\phi ,\rho\sin\phi) $
	 and write the solutions of~\eqref{app:eigenFunEq} as
	\begin{equation}\label{app:steerableEigenFun}
	\psi_{m,k}(r,\theta) = \frac{1}{\sqrt{2\pi}}R_{m,k}(r)e^{im\theta}, \;\;\;\:m\in\mathbb{Z},\;k\in \mathbb{N}.
	\end{equation}
	Substituting~\eqref{app:steerableEigenFun} into~\eqref{app:eigenFunEq} yields
	\begin{align}\label{app:eigenFuncitionEqPolar}
	\lambda_{m,k}R_{m,k}(r)e^{im\theta} &= \int_{a\mathbb{D}}K(r,\theta,r',\theta')	R_{m,k}(r')e^{im\theta'} r'dr'd\theta'.
	\end{align}
	Note that~\eqref{app:FourierOfK} in polar coordinates is given by
	\begin{align*}
	&K(r,\theta,r',\theta')=\\\nonumber
	&\frac{1}{(2\pi)^2}\int_{c\mathbb{D}}G(\rho)e^{i(\rho \cos\phi ,\rho \sin\phi)\cdot(r\cos\theta-r'\cos\theta',r\sin\theta-r'\sin\theta')}\rho d\rho d\phi=\\\nonumber
	&\frac{1}{(2\pi)^2}\int_{c\mathbb{D}}G(\rho)e^{i(\rho r \cos(\phi-\theta)-\rho r'\cos(\phi-\theta'))}\rho d\rho d\phi\nonumber,
	\end{align*}
	 thus
	\begin{equation}
    \begin{aligned}
		&\lambda_{m,k}R_{m,k}(r)e^{im\theta}\\
		&=\frac{1}{(2\pi)^2}\int_{a\mathbb{D}}\int_{c\mathbb{D}}G(\rho)e^{i(\rho r \cos(\phi-\theta)-\rho r'\cos(\phi-\theta'))} \\
        & \qquad \qquad \qquad \qquad \qquad \times R_{m,k}(r')e^{im\theta'} r' \rho dr'd\theta'd\rho d\phi\\
		&=\frac{1}{(2\pi)^2}\int_0^a\int_{c\mathbb{D}} \left(\int_{-\pi}^\pi e^{i(m\theta'-\rho r'\cos(\phi-\theta'))} d\theta'\right) G(\rho)e^{i\rho r \cos(\phi-\theta)} \\
        &\qquad \qquad \qquad \qquad \qquad  \times R_{m,k}(r') r' \rho dr'd\phi d\rho.
        \end{aligned}
        \label{eq:eigenFunNoBessel}
	\end{equation}
	Changing variables in the innermost integral in~\eqref{eq:eigenFunNoBessel} to $ \theta' = \phi - \alpha -\frac{\pi}{2}$ gives
    \begin{equation}
	\begin{aligned}
		\int_{-\pi}^\pi e^{i(m\theta'-\rho r'\cos(\phi-\theta'))} d\theta' &= e^{im(\phi-\frac{\pi}{2})}\int_{-\pi}^\pi e^{i(\rho r'\sin\alpha-m\alpha)} d\alpha\\
		&=e^{im(\phi-\frac{\pi}{2})}2\pi J_m(\rho r'),
	\end{aligned}
    \label{eq:inner integral 1}
    \end{equation}
	where $ J_m $ is the $m$'th order Bessel function of the first kind.
	Substituting back~\eqref{eq:inner integral 1} in~\eqref{eq:eigenFunNoBessel} gives
	\begin{multline}\label{eq:star star}
		\lambda_{m,k}R_{m,k}(r)e^{im\theta}=\frac{1}{2\pi}\int_0^a\int_0^c \left(\int_{-\pi}^\pi e^{i(m(\phi-\frac{\pi}{2}) +\rho r \cos(\phi-\theta))}d\phi\right) \\ \times J_m(\rho r')G(\rho) R_{m,k}(r') r' \rho dr' d\rho.
	\end{multline}
	Changing variables in the innermost integral in~\eqref{eq:star star} to $ \phi= \theta - \alpha +\frac{\pi}{2}$ gives
    \begin{equation}\label{eq:inner integral 2}
	\begin{aligned}
		\int_{-\pi}^\pi e^{i(m(\phi-\frac{\pi}{2}) +\rho r \cos(\phi-\theta))}d\phi &= e^{im\theta}\int_{-\pi}^\pi e^{i(\rho r \sin\alpha-m\alpha) }d\phi\\
		&=2\pi e^{im\theta}J_{m}(\rho r).
	\end{aligned}
    \end{equation}	
    Using~\eqref{eq:inner integral 2} in~\eqref{eq:star star} we get
	\begin{align*}
		\lambda_{m,k}R_{m,k}(r)&=\int_0^a\left(\int_0^cJ_m(\rho r)J_m(\rho r')G(\rho)r'\rho d\rho \right) R_{m,k}(r')dr' \\\nonumber
		&=\int_0^ah_m(r,r')R_{m,k}(r')dr' ,
	\end{align*}
	where $ h_m(r,r') =  \int_0^cJ_m(\rho r)J_m(\rho r')G(\rho)r'\rho d\rho $.

	\section{Estimating the solution of~\protect\eqref{eq:sFactor}}\label{app:estimating the RPSD}
	In this section, we outline an algorithm for estimating the solution of~\eqref{eq:sFactor}, that is of
    \begin{equation}\label{app:NMMF}
	\begin{gathered}
		\{\hat{\gamma},\hat{\nu},\hat{\alpha}\} = \underset{\tilde{\gamma},\tilde{\nu}, \tilde{\alpha}}{\operatorname{arg\,min}} \sum_{i=1}^M ||\mathcal{S}_i-\tilde{\nu} -\tilde{\alpha}_i \tilde{\gamma}||^2,\\
		\text{s.t.} \;\;\; \tilde{\gamma},\tilde{\nu}\in\mathbb{R}_+^l,\;\tilde{\alpha}_i\in[0,1].
	\end{gathered}
    \end{equation}
		Equation~\eqref{app:NMMF} can be viewed as an instance of a non-negative matrix factorization problem (NNMF)~\citep{gillis2017introduction}, which is NP-hard. One method for estimating solutions of NNMF problems is by alternating least-squares (ALS)~\citep{kim2008nonnegative}. The idea behind this iterative method is solving, in each step,  the NNMF minimization problem for one of the variables, while treating the others as constants. If the method converges, then it converges to a local minimum of the NNMF problem. In our case, the variables are the vectors $ \hat{\gamma},\hat{\nu},\hat{\alpha} $, and the ALS iterations are given by (for iteration $i$)
		\begin{equation}\label{app:ALS iterations}
        \begin{aligned}
			\{\hat{\alpha}^{(i)}\} &= \underset{\tilde{\alpha}}{\operatorname{arg\,min}} \sum_{k=1}^M ||\mathcal{S}_k-\hat{\nu}^{(i-1)} -\tilde{\alpha}_k \hat{\gamma}^{(i-1)}||^2,\\
			\{\hat{\gamma}^{(i)}\} &= \underset{\tilde{\gamma}}{\operatorname{arg\,min}} \sum_{k=1}^M ||\mathcal{S}_k-\hat{\nu}^{(i-1)} -\hat{\alpha}_k^{(i)} \tilde{\gamma}||^2,\\
			\{\hat{\nu}^{(i)}\} &= \underset{\tilde{\nu}}{\operatorname{arg\,min}} \sum_{k=1}^M ||\mathcal{S}_k-\tilde{\nu} -\hat{\alpha}_k^{(i)} \hat{\gamma}^{(i)}||^2,\\
			s.t&\;\;\;\tilde{\gamma},\tilde{\nu}\in\mathbb{R}_+^l,\;\tilde{\alpha}_k\in[0,1].
        \end{aligned}
		\end{equation}
		As the initial vectors $ \hat{\gamma}^{(0)}, \hat{\nu}^{(0)}$ we set
		\begin{align}\label{app: ALSinitial}
		\hat{\gamma}^{(0)}&=\mathcal{S}_{i_0}-\mathcal{S}_{j_0},\quad &\{i_0,j_0\}=\argmax_{i,j}\norm{\mathcal{S}_i-\mathcal{S}_j}_1,\\\nonumber
		\hat{\nu}^{(0)}&=\mathcal{S}_{k_0},\quad&\{k_0\}=\argmin_{i}\norm{\mathcal{S}_i}_1,
		\end{align}
		where $  \norm{u}_1 = \sum_{i=1}^{l}\lvert u_i\rvert $ for $ u\in\mathbb{R}^l $. To justify this choice, recall that $ \mathcal{S} $ is modeled as
		 \begin{equation}\label{app:analytic factorization of S}
		 \mathcal{S}_i = \nu + \alpha_i\gamma,\quad 1\leq i\leq M.
		 \end{equation}
		 thus,
		\begin{align*}
			\{i_0,j_0\}&=\argmax_{i,j}\norm{\mathcal{S}_i-\mathcal{S}_j}_1=\argmax_{i,j}\norm{\left(\alpha_i-\alpha_j\right)\gamma}_1\\\nonumber
			&=\argmax_{i,j}|\alpha_i-\alpha_j|\norm{\gamma}_1
		\end{align*}
		which implies
		\begin{equation*}
			\hat{\gamma}^{(0)} = \left(\alpha_{\text{max}}-\alpha_{\text{min}}\right)\cdot\gamma.
		\end{equation*}
		Since we expect at least one patch to include a complete particle, and at least one patch to contain only noise, we get $ \alpha_{\text{max}}\approx1,\;\alpha_{\text{min}}\approx0 $, which yields
		\begin{equation*}
			\hat{\gamma}^{(0)} = \left(\alpha_{\text{max}}-\alpha_{\text{min}}\right)\cdot\gamma\approx (1-0)\cdot\gamma=\gamma.
		\end{equation*}
		That is $ \hat{\gamma}^{(0)} $ is initialized close to its true value.
		As for $ \hat{\nu}^{(0)}$,~\eqref{app:analytic factorization of S} implies
		\begin{align*}
				\{k_0\}&=\argmin_k\norm{\mathcal{S}_k}_1=\argmin_{k}\norm{\nu + \alpha_k\gamma}_1\\\nonumber
				&=\argmin_{k}\left(\norm{\nu}_1+\alpha_k\norm{\gamma}_1\right)=\argmin_{k}\alpha_k.
		\end{align*}	
		where the third equality holds due to the positivity of all arguments. As written above, $ \alpha_{\text{min}}\approx 0 $ and hence
		\begin{equation*}
		\hat{\nu}^{(0)} \approx \nu+0\cdot\gamma = \nu,
		\end{equation*}
		that is, $ \hat{\nu}^{(0)} $ is also initialized close to its true value.
		
		Next, we derive an explicit formula for  the solution of~\eqref{app:ALS iterations}.  Using basic algebra, we can rewrite the sum in ~\eqref{app:NMMF} in three equivalent forms
		\begin{align}\label{app:ALS sum}
		&\sum_{i=1}^M \norm{\mathcal{S}_i-\tilde{\nu} -\tilde{\alpha}_i \tilde{\gamma}}^2=\\\nonumber
		&\sum_{k=1}^{l}\left[\left(\sum_{i=1}^{M}\tilde{\alpha}_i^2\right)\tilde{\gamma}_k^2-2\left(\sum_{i=1}^{M}\tilde{\alpha}_i\tilde{\nu}_k+\tilde{\alpha}_i\mathcal{S}_{ki}\right)\tilde{\gamma}_k+\sum_{i=1}^{M}\left(\mathcal{S}_{ki}-\tilde{\nu}_k\right)^2\right]=\\\nonumber
		&\sum_{k=1}^{l}\left[M\tilde{\nu}_k^2-2\left(\sum_{i=1}^{M}\tilde{\alpha}_i\tilde{\gamma}_k+\mathcal{S}_{ki}\right)\tilde{\nu}_k+\sum_{i=1}^{M}\left(\mathcal{S}_{ki}-\tilde{\alpha}_i\tilde{\gamma}_k\right)\right]=\\\nonumber
		&\sum_{i=1}^{M}\left[\left(\sum_{k=1}^{l}\tilde{\gamma}_k^2\right)\tilde{\alpha}_i^2-2\left(\sum_{k=1}^{l}\tilde{\nu}_k\tilde{\gamma}_k+\tilde{\gamma}_k\mathcal{S}_{ki}\right)\tilde{\alpha}_i+\sum_{k=1}^{l}\left(\mathcal{S}_{ki}-\tilde{\nu}_k\right)^2\right].
		\end{align}
		Treating $\tilde{\gamma}$ as a variable and $ \tilde{\nu},\tilde{\alpha} $ as constants in the second line, gives a sum of convex parabolas for the variables $ \{\tilde{\gamma}_k\}_{k=1}^l $. The minimum of this sum, under the constraint $\tilde{\gamma}_k \geq 0$ is attained at the minimum of these parabolas whenever they are positive and at $ \tilde{\gamma}_k =0$ else. The same applies when treating  $\tilde{\nu}$ as the variable in the third line (and the other as constants). Treating $ \tilde{\alpha} $ as the variable in the forth line, implies that the minimum is given as follow. For each parabola in the sum~\eqref{app:ALS sum}, if its minimum is obtained at a negative $ \tilde{\alpha}_k $, then the minimum of the sum is attained at $ \tilde{\alpha}_k = 0 $. If the minimum is obtained at $ \tilde{\alpha}_k > 1 $, we set $\tilde{\alpha}_k=1$. Otherwise, the minimum of the sum is attained at the minimum of the parabola. To conclude, we get the following formulas for the solutions of the $ i $'th iteration in~\eqref{app:ALS iterations}
		\begin{equation}\label{app: ALSest}
			\begin{alignedat}{3}
			\hat{\alpha}_j^{(i)} &=\left[\frac{\sum_{k=1}^{l}\hat{\gamma}_k^{(i-1)}\cdot\left(\mathcal{S}_{k,j}-\hat{\nu}_k^{(i-1)}\right)}{\sum_{k=1}^{l}\hat{\gamma}_k^{(i-1)}}\right]_{+,\leq 1}, &\quad &1\leq j\leq M,\\
			\hat{\gamma}_j^{(i)} &=\left[\frac{\sum_{k=1}^{M}\hat{\alpha}_k^{(i)}\cdot\left(\mathcal{S}_{j,k}-\hat{\nu}_j^{(i-1)}\right)}{\sum_{k=1}^{M}\hat{\alpha}_k^{(i)}}\right]_{+}, &\quad  &1\leq j\leq l,\\
			\hat{\nu}_j^{(i)} &=\left[\frac{\sum_{k=1}^{M}\mathcal{S}_{j,k}-\hat{\alpha}_k^{(i)}\hat{\gamma}_j^{(i)}}{M}\right]_{+}, &\quad& 1\leq j\leq l,
		\end{alignedat}
		\end{equation}
		where for $ r\in\mathbb{R} $ \[
		\left[r\right]_{+,\leq 1} =
		\begin{cases}
		0&\quad r<0,\\
		r &\quad 0\leq r \leq 1,\\
		1 &\quad r>1,\\
		\end{cases}
		\]
		and
		\[
		\left[r\right]_{+} =
		\begin{cases}
		0&\quad r<0,\\
		r &\quad r\geq0.\\
		\end{cases}
		\]
		One can prove that  if $ \mathcal{S} $ can be factorized as in ~\eqref{app:analytic factorization of S}, and the iterations of~\eqref{app: ALSest} converge, then in the limit of $ i \to \infty $ in~\eqref{app: ALSest}, it holds that
		\begin{align}\label{app:ALS limits}
		\hat{\gamma} = a\gamma,\;\; \hat{\alpha}_{k} = \frac{1}{a}\left(\alpha_k + b\right),\;\; \hat{\nu}= \nu -b\gamma,
		\end{align}
		where $ a,b \in \mathbb{R} $ can be estimated from the noise variance. This implies that we can recover $ \gamma,\alpha,\nu $. We omit the details of the proof.  In~\ref{app:noise varience estimation}, we describe a method to estimate the noise variance from a given micrograph. The procedure for solving~\eqref{app:NMMF} is summarized in Algorithm~\ref{alg: ALS}.
			\begin{algorithm}
			\caption{\textbf{ALS algorithm}}
			\label{alg: ALS}
			\begin{algorithmic} [1]
				\Statex{\textbf{Required:} Matrix $ \mathcal{S} $.}
				\State 	 Compute the initial vectors $ \hat{\nu}^{(0)}, \hat{\gamma}^{(0)}$ using~\eqref{app: ALSinitial} and $ \hat{\alpha}^{(0)}, \hat{\alpha}^{(1)}, \hat{\nu}^{(1)}, \hat{\gamma}^{(1)} $ using~\eqref{app: ALSest}.
				\State $ i=1 $
				\While{$\norm{\hat{\nu}^{(i)}-\hat{\nu}^{(i-1)}},\norm{\hat{\gamma}^{(i)}-\hat{\gamma}^{(i-1)}},\norm{\hat{\alpha}^{(i)}-\hat{\alpha}^{(i-1)}}\geq \epsilon$ }
				\State Compute $ \hat{\nu}^{(i+1)}, \hat{\gamma}^{(i+1)},\hat{\alpha}^{(i+1)} $ using~\eqref{app: ALSest}
				\State$  i = i+1 $
				\EndWhile
				\Statex{\textbf{Output:} $\hat{\nu}^{(i)},\hat{\gamma}^{(i)},\hat{\alpha}^{(i)}$}
			\end{algorithmic}
		\end{algorithm}
		
		Finally, we return to the computation of the columns of~$\mathcal{S}$. Each column is the RPSD of a micrograph patch of size $ l \times l$ with stride $ d $, as described at the beginning of Section~\ref{subsec:psdEstimation}. These parameters are determined as follows. In~\eqref{app:analytic factorization of S} we assume that $ \mathcal{S} $ can be factorized as a combination  of only $ \nu,\gamma,\alpha $. However, this model is only an approximation, and in practice it holds that
		\begin{equation}\label{eq:S with error term}
		\mathcal{S} = \nu+\alpha^T\gamma+\mathcal{E},
		\end{equation}
		where $ \mathcal{E} $ is an error term matrix. Under the assumptions that the expectancy of the rows of $ \mathcal{E} $ is zero, and that the columns of $ \mathcal{E} $ are uncorrelated and with the same variance, one can prove that the limit of~\eqref{app: ALSest} when using $\mathcal{S}$ of~\eqref{eq:S with error term} stays the same as in~\eqref{app:ALS limits}, but with different values for $ a,b $ (we omit the proof). These values can be estimated from the noise variance, and therefore, we can recover $ \gamma,\alpha, \nu $ for the more general model~\eqref{eq:S with error term}. Now, taking $ d<l $ clearly breaks the assumptions for $ \mathcal{E} $ and taking $ d>l $ does not use all of the data, hence we used $ d=l $. As for the parameter~$l$, ideally we would like a patch to contain no more then one complete particle. Taking $ l $ much bigger then the particle diameter can cause patches to contain two or more particles, and taking it much smaller will not allow patches to contain enough of the particle. In practice, it seems that taking $ l $ equal 80\% of the particle diameter yields good results.
		\section{Estimating the noise variance}\label{app:noise varience estimation}
		The noise variance is needed to recover $ \gamma,\alpha,\nu $ from~\eqref{app:ALS limits} and to evaluate~\eqref{eq:covMat}.
		Assuming that both the noise and the particle function are generated from a wide sense stationary random process and that they are uncorrelated, it follows that the variance of a given micrograph patch is the sum of the noise variance and a fraction of the particle's variance, which is proportional to the fraction of the particle contained in the patch.  As the variance is a positive number, patches with lower variance are more likely to contain only noise. With that in mind, we partition the micrograph into overlapping particle-size patches with stride equal to $ 1 $. Then, we approximate the variance of each patch as explained in~\cite{wasserman2013all} and finally estimate the noise variance by averaging $ x\% $ of the lowest variance results. Choosing $ x $ depends on the number of particles in the micrograph, which is unknown to us. We have observed experimentally that taking $ x = 25\% $ provides satisfactory results in low-SNR conditions.

		\section{Deriving the approximation~\eqref{eq:cleanCovApprox}}\label{app:LRT derivation}
		In this section, we derive an approximation for the particle's covariance matrix $\mathbb{E}\left[\mathcal{P}\mathcal{P}^T\right]$, where \begin{equation}\label{app:connection between the function and samples}
			\mathcal{P}_{i} = f(i),\quad i \in \mathcal{U}
		\end{equation}
(see also Section~\ref{subsec:particleDetection}).
		The Karhunen Loeve Theorem~\citep{maccone2009simple} states that
		\begin{align*}
			f(x) &= \sum_{i=1}^{\infty}z_i\psi_i(x),
		\end{align*}
		where $ \psi_i $ are given by~\eqref{eq:templatesDef}, and moreover, $\mathbb{E}\left[z_iz_j\right]=\delta_{i,j}\lambda_i$. Therefore, the autocorrelation $K(x,y) = \mathbb{E}\left[f(x)f(y)\right]$ can be written as
		\begin{align*}
			\mathbb{E}\left[f(x)f(y)\right] &= \mathbb{E}\left[\left(\sum_{i=1}^{\infty}z_i\psi_i(x)\right)\left(\sum_{j=1}^{\infty}z_j\psi_j(y)\right)\right]\\\nonumber
			&=\sum_{i=1}^{\infty}\sum_{j=1}^{\infty}\psi_i(x)\psi_j(y)\mathbb{E}\left[z_iz_j\right]\\\nonumber
			&=\sum_{i=1}^{\infty}\sum_{j=1}^{\infty}\psi_i(x)\psi_j(y)\delta_{i,j}\lambda_i\\\nonumber
			&=\sum_{i=1}^{\infty}\psi_i(x)\psi_i(y)\lambda_i.
		\end{align*}
		Using~\eqref{app:connection between the function and samples} we get
		\begin{align}\label{eq:truncate autocorr}
		\mathbb{E}\left[\mathcal{P}\mathcal{P}^T\right]_{(i,j)} = 	\mathbb{E}\left[\mathcal{P}_i\mathcal{P}_j\right]=\mathbb{E}\left[f(i)f(j)\right]= \sum_{k=1}^{\infty}\psi_k(i)\psi_k(j)\lambda_k.
		\end{align}
		In order to define a truncation rule for the infinite sum in~\eqref{eq:truncate autocorr}, we consider the relative error term
		\begin{align}
		\frac{\left\Vert \sum_{k=1}^{\infty}\psi_k\psi_k\lambda_k-\sum_{k=1}^{N}\psi_k\psi_k\lambda_k\right\Vert}{\left\Vert \sum_{k=1}^{\infty}\psi_k\psi_k\lambda_k\right\Vert}&=\frac{\left\Vert\sum_{k=N+1}^{\infty}\psi_k\psi_k\lambda_k\right\Vert}{\left\Vert\sum_{k=1}^{\infty}\psi_k\psi_k\lambda_k\right\Vert} \label{app:rel err 1}\\
		&=\frac{\sum_{k=N+1}^{\infty}\lambda_k}{\sum_{k=1}^{\infty}\lambda_k}, \label{app:rel err 2}
		\end{align}
		where $ ||\cdot|| $  is the $ L_2\left(a\mathbb{D}\times a\mathbb{D}\right) $ standard norm and~\eqref{app:rel err 2} follows from~\eqref{app:rel err 1} due to orthogonality of $ \psi_k $. Then, we define the following truncation rule. First we truncate the infinite sum at $ M $ terms where $ M $ is chosen such that
		\begin{equation*}
			\frac{\lambda_M - \lambda_{M+1}}{\lambda_M} \leq 10^{-6},
		\end{equation*}
		 and then, we truncate it again to $ N $ terms such that
		 \begin{equation}\label{app:truncation rule}
			\frac{\sum_{k=N+1}^{M}\lambda_k}{\sum_{k=1}^{M}\lambda_k} \leq 0.01.
		\end{equation}
		By standard linear algebra, it can be shown that the truncation rule of~\eqref{app:truncation rule} yields the desired approximation of~\eqref{eq:cleanCovApprox}
		\begin{equation}
		\mathbb{E}\left[\mathcal{P}\mathcal{P}^T\right] = \sum_{k=1}^{\infty}\psi_k(i)\psi_k(j)\lambda_k \approx \sum_{k=1}^{N}\psi_k(i)\psi_k(j)\lambda_k = \Psi\Lambda\Psi^T
		\end{equation}
		where $ \Psi $ and $ \Lambda $ are defined in Section~\ref{subsec:particleDetection}
		\section{Derivation and computation of~\eqref{eq:scoringMatrix}}
			 Our goal is to compute the score
			 \begin{equation}\label{app:score test 1}
			 	\text{score}(i,j):=\log\left(\frac{p(\mathcal{I}^{i,j}|H_1)}{p(\mathcal{I}^{i,j}|H_0)}\right).
			 \end{equation}
			 Under the assumptions that the particle is distributed multivariate normal, and that the noise is white, the probability density function (PDF) of a micrograph patch containing a particle plus noise is
			 \begin{equation}\label{app: denisity of particle with added noise}
			 p\left(\mathcal{I}|H_1\right) = \frac{1}{\sqrt{|2\pi\Sigma|}}e^{-\frac{1}{2}\mathcal{I}^T\Sigma^{-1}\mathcal{I}},
			 \end{equation}
			 where $ |\cdot| $ is a matrix determinant. The PDF of a patch with noise only is
			\begin{equation}\label{app: denisity of particle with noise}
			p\left(\mathcal{I}|H_0\right) = \frac{1}{\sqrt{(2\pi\sigma^{2})^{\lfloor2a+1\rfloor^2}}}e^{-\frac{1}{2\sigma^2}\mathcal{I}^T\mathcal{I}}.
			\end{equation}
			Substituting the above PDFs in ~\eqref{app:score test 1} and replacing the random variable $ \mathcal{I} $ with its $ (i,j) $ patch realization $\mathcal{I}^{i,j}$  gives (see also~\eqref{eq:scoringMatrix})
			\begin{align}\label{app:score test 2} \text{score}(i,j)=\frac{1}{2}\left(\log\left(\frac{|\Sigma|}{\sigma^{2\lfloor2a+1\rfloor^2}}\right)+\left(\mathcal{I}^{i,j}\right)^T\left(\sigma^{-2}I-\Sigma^{-1}\right)\mathcal{I}^{i,j}\right).
			\end{align}
				We estimate $ \Sigma $ for~\eqref{app:score test 2} using the approximation~\eqref{eq:cleanCovApprox} for the particle's covariance matrix, that is,
				\begin{equation*}
				\Sigma = \mathbb{E}\left[\mathcal{P}\mathcal{P}^T\right]+ \sigma^2I \approx\Psi\Lambda\Psi^T + \sigma^2I :=\hat{\Sigma}.
				\end{equation*}
				 Computing $ \hat{\Sigma}^{-1} $ and $ |\hat{\Sigma}| $ via the singular values decomposition (SVD)~\citep{suli2003introduction} takes $O(\lfloor 2a+1 \rfloor^2 N^2)$ operations per micrograph, where $ N $ is  given by the truncation rule of~\eqref{app:truncation rule}. The computational cost of directly computing
				\begin{equation}\label{app:per patch mull}
					P_{i,j} = \mathcal{I}_{i,j}^T\left(\sigma^{-2}I-\hat{\Sigma}^{-1}\right)\mathcal{I}_{i,j} ,
				\end{equation}
				is $ O(\lfloor 2a+1 \rfloor^6)$ operations per patch. In order to reduce the latter complexity, consider the approximate covariance matrix
			\begin{equation}\label{app:cov matrix}
				\hat{\Sigma} =  \Psi\Lambda\Psi^T + \sigma^2I,
			\end{equation}
			and factorize it using the QR decomposition~\citep{suli2003introduction} as
			\begin{equation}\label{app:qr}
				\Psi = QR,
			\end{equation}
			where $ Q\in \mathbb{R}^{\lfloor 2a+1 \rfloor^2 \times \lfloor 2a+1 \rfloor^2}$ is an orthogonal matrix and $ R\in\mathbb{R}^{\lfloor 2a+1 \rfloor^2 \times N}$ is an upper triangular matrix whose last $ \lfloor 2a+1 \rfloor^2 - N $ rows are zero. We denote $ R = \begin{pmatrix}	\tilde{R} \\ 0  \end{pmatrix} $ where $ \tilde{R} \in\mathbb{R}^{N\times N} $, and substitute~\eqref{app:qr} in~\eqref{app:cov matrix}, resulting in
			\begin{equation*}
			\hat{\Sigma} = Q\left(R\Lambda R^T + \sigma^2I\right)Q^T.
			\end{equation*}
			Note that $ R\Lambda R^T + \sigma^2I$ is a block diagonal matrix of the form
			\begin{equation*}
			R\Lambda R^T + \sigma^2I = \begin{pmatrix} H & 0\\0&\sigma^2I_2 \end{pmatrix},
			\end{equation*}
			where $ H = \tilde{R}\Lambda\tilde{R}^T + \sigma^2I_1  $ is of size $ N\times N $, $ I_1 $ is the Identity matrix of the same size as $ H $, and $ I_2$ is the identity matrix of size $ (\lfloor 2a+1\rfloor^2 -N)\times (\lfloor 2a+1\rfloor^2 -N)$.
			The determinant of $ \hat{\Sigma} $ is
			\begin{equation}\label{app:sigma determinant}
			|\hat{\Sigma}| = |H|\sigma^{\lfloor 2a+1 \rfloor^2 - N },
			\end{equation}
		    and its inverse is
		    \begin{equation*}
		    \hat{\Sigma}^{-1} = Q\begin{pmatrix}
		    H^{-1}&0\\0&\sigma^{-2}I_2
		    \end{pmatrix} Q^T.
		    \end{equation*}
			In addition,
			\begin{align*}
			\sigma^{-2}I-\hat{\Sigma}^{-1}&=\sigma^{-2}\begin{pmatrix}
			I_1&0\\0&I_2\end{pmatrix}-Q\begin{pmatrix} H^{-1}&0\\0&\sigma^{-2}I_2\end{pmatrix}Q^T\\\nonumber
			&=Q\begin{pmatrix}\sigma^{-2}I_1 -H^{-1}&0\\0&0\end{pmatrix}Q^T.
			\end{align*}
			Substituting the latter back in~\eqref{app:per patch mull} yields
            \begin{equation}\label{app:inner product not simplified}
			\begin{aligned}
			\mathcal{I}^T\left(\sigma^{-2}I-\hat{\Sigma}^{-1}\right)\mathcal{I} &=\mathcal{I}^TQ\begin{pmatrix}\sigma^{-2}I_1 -H^{-1}&0\\0&0\end{pmatrix}Q^T\mathcal{I}\\
			&=\left(Q^T\mathcal{I}\right)^T\begin{pmatrix}\sigma^{-2}I_1 -H^{-1}&0\\0&0\end{pmatrix}Q^T\mathcal{I}.
			\end{aligned}
            \end{equation}
			Denote the first $N$  coordinates of the vector $ Q^T\mathcal{I} $ by $ v $. Then, the last expression is simplified to
			\begin{equation}\label{app:inner product simplifief}
			\mathcal{I}^T\left(\sigma^{-2}I-\hat{\Sigma}^{-1}\right)\mathcal{I} = \sigma^{-2}v^Tv-v^TH^{-1}v.
			\end{equation}
			Substituting~\eqref{app:sigma determinant} and~\eqref{app:inner product simplifief} in~\eqref{app:score test 2} and changing  $ \mathcal{I} $ to  $ \mathcal{I}_{i,j} $ gives
			\begin{equation}\label{app:simplified acore test}
			\text{score}(i,j)=\frac{1}{2}\left(log\left(\frac{\sigma^{2N}}{|H|}\right) + \sigma^{-2}v^Tv-v^TH^{-1}v\right).
			\end{equation}
			The complexity of computing  $ H $ via the QR decomposition of $ \hat{\Sigma} $  is $ O\left(\lfloor2a+1\rfloor^2 N^2\right) $ operations. Then, computing  $ H^{-1} $ and $ |H| $ via $ SVD $ requires $ O(N^3) $ operations, both computed per micrograph. Computing $ v^TH^{-1}v $ requires $ O(N^3) $ operations, and computing $ v^Tv $ requires $ O(N^2) $ operations, both computed per patch. Finally, in order to compute the vector $ v $, we need to compute only the first $ N $ coordinates of $ Q^T\mathcal{I} $. This can be implemented using $O\left(\lfloor2a+1\rfloor^2N\right) $ operations per patch. To conclude, the per patch computational complexity has been reduced from $ O(\lfloor2a+1\rfloor^6) $ to $ O\left(\lfloor2a+1\rfloor^2N+N^3\right) $. In practice, this speeds up the algorithm by three orders of magnitude.
%\bibliographystyle{abbrvnat}
%\setcitepstyle{authoryear}
\bibliography{refYoel4}

	\end{document}